\documentclass[usenatbib]{mn2e}

\usepackage{amsmath}
\usepackage{natbib}
\usepackage{epsfig}
\usepackage{txfonts}

\newcommand{\matpol}[4]{\left(\begin{array}{cc} #1 & #2 \\ #3 & #4 \end{array}\right)}

\newcommand{\Mpc}{{\rm Mpc}}

\newcommand{\expf}[1]{{{\rm e}^{#1}}}

\newcommand{\zmudc}{{z_{\rm dc}}}

\newcommand{\nbb}{{n^{\rm pl}}}

\newcommand{\nS}{n_{\rm S}}
\newcommand{\nT}{n_{\rm T}}

\newcommand{\taudot}{\dot{\tau}}
\newcommand{\kD}{k_{\rm D}}
\newcommand{\rs}{r_{\rm s}}
\newcommand{\cs}{c_{\rm s}}

\newcommand{\id}{{\,\rm d}}

\newcommand{\beq}{\begin{equation}}   %

\newcommand{\eeq}{\end{equation}}   %

\newcommand{\beqa}{\begin{eqnarray}}   %

\newcommand{\eeqa}{\end{eqnarray}}   %

\newcommand{\beal}{\begin{align}}

\newcommand{\enal}{\end{align}}

\newcommand{\bspl}{\begin{split}}

\newcommand{\espl}{\end{split}}

\newcommand{\bsub}{\begin{subequations}}

\newcommand{\esub}{\end{subequations}}

\newcommand{\bmulti}{\begin{multline}}   %

\newcommand{\beqm}{\begin{mathletters}}   %

\newcommand{\eeqm}{\end{mathletters}}   %

\newcommand{\Ne}{N_{\rm e}}

\newcommand{\sigT}{\sigma_{\rm T}}

\newcommand{\vek} [1]{\mbox{\boldmath${#1}$\unboldmath}}

\newcommand{\pot}[2]{#1 \times 10^{#2}}

\usepackage{color}

\usepackage{hyperref}
\usepackage{grffile}
\usepackage{graphics}
\usepackage{booktabs}

\title[Dissipation of tensor modes]
{Spectral distortions from the dissipation of tensor perturbations}

\author[Chluba et al.]{
 Jens Chluba$^{1}$\thanks{E-mail: jchluba@pha.jhu.edu}, 
 Liang Dai$^{1}$\thanks{E-mail: ldai@pha.jhu.edu}, 
 Daniel Grin$^{2}$,
 Mustafa A. Amin$^{3}$ and 
 Marc Kamionkowski$^{1}$
 \\
$^{1}$ Department of Physics and Astronomy, Johns Hopkins University, 
3400 N. Charles St., Baltimore, MD 21218, USA
\\
$^{2}$ Department of Astronomy and Astrophysics, University of Chicago, Chicago, Illinois 60637, USA
\\
$^{3}$ Kavli Institute for Cosmology at Cambridge, Madingley Rd, Cambridge CB3 OHA, UK
}

\date{{\vspace{-10.0mm}Received 2014 July 14.}}

\begin{document}

\maketitle

\begin{abstract}
Spectral distortions of the cosmic microwave background (CMB) may become a powerful probe of primordial perturbations at small scales. Existing studies of spectral distortions focus almost exclusively on primordial scalar metric perturbations. Similarly, vector and tensor perturbations should source CMB spectral distortions. In this paper, we give general expressions for the effective heating rate caused by these types of perturbations, including previously neglected contributions from polarization states and higher multipoles. We then focus our discussion on the dissipation of tensors, showing that for nearly scale invariant tensor power spectra, the overall distortion is some six orders of magnitudes smaller than from the damping of adiabatic scalar modes. We find simple analytic expressions describing the effective heating rate from tensors using a quasi-tight coupling approximation. In contrast to adiabatic modes, tensors cause heating without additional photon diffusion and thus over a wider range of scales, as recently pointed out by \citet{Ota2014}. Our results are in broad agreement with their conclusions, but we find that small-scale modes beyond $k\simeq \pot{2}{4}\,\Mpc^{-1}$ cannot be neglected, leading to a larger distortion, especially for very blue tensor power spectra. At small scales, also the effect of neutrino damping on the tensor amplitude needs to be included.
\end{abstract}



\section{Introduction}
\label{sec:intro}
Tiny deviations of the cosmic microwave background (CMB) spectrum from a perfect blackbody -- commonly referred to as spectral distortions -- provide a powerful tool for studying the thermal history of our Universe \citep[see][for overview]{Chluba2011therm, Chluba2013fore}. In particular, the possibility of probing the primordial power spectrum of curvature perturbations at very small scales (wavenumbers $3\,\Mpc^{-1}\lesssim k \lesssim \pot{2}{4}\,\Mpc^{-1}$) using CMB spectral distortions \citep[e.g.,][]{Sunyaev1970diss, Daly1991, Hu1994, Chluba2012, Pajer2012} has stimulated an increased interest in this topic. 

For scalar perturbations, the distortion depends on the amplitude and shape of the power spectrum at small scales \citep{Chluba2012, Chluba2012inflaton, Chluba2013PCA}. It also matters whether adiabatic or isocurvature modes are initially excited \citep{Hu1994isocurv, Dent2012, Chluba2013iso}. Spectral distortions could furthermore be used to probe scale-dependent non-Gaussianity in the ultra-squeezed limit through spatial variations of $\mu$-distortions at large scales \citep{Pajer2012, Ganc2012, Biagetti2013} and constrain the energy scale of phase transitions in the early universe \citep{Amin2014}. Spectral distortions are thus an invaluable new source of information about early-universe physics, which is complementary and independent of the directly observable CMB temperature and polarization anisotropies at larger angular scales.

The distortion from primordial perturbations is created because the superposition of blackbodies at different temperatures is not itself a blackbody \citep{Zeldovich1972, Chluba2004, Stebbins2007}. A spatially varying photon field at different scales is set up by inflation and the mixing of blackbodies is accomplished by Thomson scattering. This dissipation process sources an average $y$-distortion, also known in connection with the thermal Sunyaev-Zeldovich effect \citep{Zeldovich1969}. The $y$-distortion then slowly evolves into a $\mu$-distortion by Compton scattering and, if there was enough time, may fully thermalize with the help of double Compton and bremsstrahlung emission, depending on when the mixing occurred \citep[e.g., see][]{Hu1993}. 

The photon quadrupole anisotropy plays a crucial role in the dissipation process, giving rise to shear viscosity in the photon fluid \citep{Weinberg1971, Kaiser1983}. For the dissipation it is, however, irrelevant which process creates the quadrupole anisotropy.  For scalar perturbations, fluctuations in the photon temperature are sourced mainly in the local monopole through the Sachs-Wolfe effect. Photon diffusion, free streaming and bulk flows further source dipole, quadrupole and higher multipoles of the photon field. There is no direct source of quadrupole anisotropies from scalar perturbations and the photon diffusion process controls its amplitude and hence the dissipation rate, which is most effective around the dissipation scale, $\kD$ \citep[see][for more discussion of the physics]{Hu1995CMBanalytic}. In the pre-recombination era ($z\gtrsim 10^4$), any fluctuation in the CMB temperature caused by scalars is erased by photon diffusion \citep[also referred to as Silk damping;][]{Silk1968}, well before it can reach the free streaming regime.

It is well known, that CMB polarization anisotropies are also sourced through the local quadrupole anisotropy \citep[e.g.,][]{BE1984}. Two types of patterns, known as curl-free $E$-modes and divergence-free $B$-modes, can be created \citep[e.g.,][]{Kamionkowski1997, Seljak1997, Kamionkowski1998}. At first order in perturbation theory, scalars only source $E$-mode patterns, while $B$-modes are indicative of tensor perturbations caused by a gravitational wave background. It is thus clear that tensor perturbations provide another contribution to the local quadrupole that differs from the one sourced by scalars. In particular, tensor modes {\it directly} give rise to a quadrupole anisotropy without the need of photon diffusion \citep[e.g.,][]{Hu1997}. Thomson scattering then mixes photons causing nearly scale independent dissipation, as also explained by \citet{Ota2014}. 

In this paper, we analyze the dissipation of tensor perturbations in the photon fluid in more detail, developing the essential physical elements of this process and showing that simple analytic expressions can be found for the effective heating rate. The recent detection of $B$-mode polarization patterns at degree angular scales by the BICEP2 team may indicate an unexpectedly large tensor to scalar ratio $r\simeq 0.2$ \citep{BICEP2results}, so that the possibility of spectral distortions from tensors merits careful consideration. While it is still open how much of this signal is truly primordial, this result seems to be in tension with the upper limits on $r$ derived indirectly from the CMB temperature power spectrum measurements of Planck \citep{Planck2013params}. Several solutions for this apparent tension have been discussed \citep[e.g.,][]{Zhang2014, Bonvin2014, Dvorkin2014, Lizarraga2014, Moss2014, Chluba2014r}. One possibility could be a strongly blue-tilted tensor power spectrum with tensor spectral index $\nT\simeq 1$ \citep[e.g.,][]{Brandenberger2014, Gerbino2014}. Although for the simplest inflation scenario, one expects $\nT\simeq - r / 8 \simeq 0$ \citep{Grishchuk1975, Starobinsky1979}, several non-standard models can accommodate $\nT\simeq \mathcal{O}(1)$ \citep[e.g.,][]{Brustein1995, Khoury2001, Boyle2004, Endlich2013}. It is thus interesting to ask what we could learn about tensors from measurements of CMB spectral distortions. Furthermore, it is important to quantify how much the dissipation of tensors could add to the distortion signal of adiabatic scalar modes.

Future constrains on the tensor power spectrum from CMB spectral distortions have recently been discussed by \citet{Ota2014}. It was shown that the damping of tensor perturbations typically causes much smaller distortions than scalar perturbations unless a very blue tensor power spectrum is assumed. Our calculations generally agree with this finding. However, we obtain a larger distortion for very blue power spectra. The main reason is that \citet{Ota2014} only included modes at $k\lesssim \pot{2}{4}\,\Mpc^{-1}(\equiv$ diffusion scale around the thermalization redshift $z\simeq \pot{2}{6}$). While this is sufficient for scalars, the damping of tensors is efficient to much smaller scales. The main difference is that gravity waves directly source a quadrupole anisotropy and no intermediate photon diffusion is required, making the damping process nearly scale independent. 
Another reason is that the total energy extracted from tensors through damping in the photon fluid is only a tiny correction to their power. This means that tensors continue to source temperature fluctuations at basically all scales and dissipation is effective even in the quasi-free streaming regime at scales smaller than the photon mean free path. Spectral distortions thus probe tensor perturbations to much smaller scales than for scalar perturbations. However, given that for the simplest inflation models $\nT\simeq 0$, overall the expected distortion signal caused by tensor perturbations remains subdominant and provides only a tiny correction to the signal from small-scale adiabatic perturbations.
We also discuss the contribution from higher multipoles and polarization showing that they only add a small correction. Dissipation of tensors in the post-recombination era is furthermore found to be subdominant.

The paper is organized as follows: in Sect.~\ref{sec:sup_basic}, we give simple expressions for the average distortion caused by the superposition of linearly polarized blackbodies of different temperatures. We then use these expressions to derive the effective heating rate for different types of perturbations (Sect.~\ref{sec:heating_rate} and \ref{sec:heating_general}). Our formulation of the problem uses the notation of \citet{Hu1997}. In Sect.~\ref{sec:heating_general}, we specialize to the case of tensor perturbations. The effective heating rates and $\mu$-distortion amplitude are discussed in Sect.~\ref{sec:mu_res}. Our main results are presented in Fig.~\ref{fig:heat_rate_examples} and \ref{fig:mu_plot}. In Sect.~\ref{sec:mu_res}, we also directly compare with \citet{Ota2014}, showing that for $\nT\simeq 1$ the distortion is underestimated by a factor of $\simeq 7$ due to additional contributions from very small scales ($k\gtrsim\pot{2}{4}\,\Mpc^{-1}$). Furthermore, the damping of tensors by the free streaming of neutrinos was neglected previously, an effect that reduces tensor power at small scales $\simeq 1.5$ times. We conclude in Sect.~\ref{sec:conclusions}.

We also included an extensive set of Appendices, in which we explicitly derive expressions describing the superposition of partially polarized blackbody radiation (Appendix~\ref{app:sup_gen_deriv}) and show that at second order in perturbation theory spectral distortions are only sourced by scattering terms, even for metric vector and tensor perturbations (Appendix~\ref{app:proof}). In Appendix~\ref{app:ana_sol_env}, we furthermore derive approximate solutions to the photon transfer functions, which capture all the phases of the evolution very well.

\vspace{-4mm}
\section{Superposition of linearly polarized blackbodies at different temperatures}
\label{sec:sup_basic}
The superposition of  blackbodies at different temperatures for unpolarized light is known to cause a $y$-type distortion at second order in the temperature difference \citep{Zeldovich1972, Chluba2004, Stebbins2007}. This just follows from a Taylor series expansion of a blackbody occupation number, $\nbb(x)=(\expf{x}-1)^{-1}$ with $x=h\nu/kT$, around reference temperature, $T_\ast\neq T$: 
\beal
\label{eq:sup_bb}
\nbb(x)&\approx  \nbb(x_\ast) + G(x_\ast) (\Theta_\ast + \Theta^2_\ast) + \frac{1}{2} Y_{\rm SZ}(x_\ast) \,\Theta^2_\ast,
\end{align}
where $x_\ast=x \, T/T_\ast=h\nu/kT_\ast$ and $\Theta_\ast=(T-T_\ast)/T_\ast$. Here, the spectrum of a temperature shift at lowest order in $\Theta$ is given by $G(x)=-x \partial_x \nbb(x)=x\expf{x}/(\expf{x}-1)^2$ and $Y_{\rm SZ}=G(x) [x \coth(x/2)-4]$ denotes a $y$-distortion \citep{Zeldovich1969}. 

We can generalize this expression to partially polarized light (linear polarization only) using a density matrix representation for the individual polarization states (see Appendix~\ref{app:sup_gen_deriv}). 
Since $Q$ and $U$ depend on the choice of the polarization basis, it is more convenient to use the combinations ${\boldsymbol M}_\pm=({\boldsymbol\sigma}_3\mp i{\boldsymbol\sigma}_1)/2$ to represent the polarization state of the system \citep[e.g., see][]{Hu1997}. With $\Theta_\pm=\Theta_Q\pm i\Theta_U$, from Eq.~\eqref{eq:occ_IQU} we find
\beal
\mathcal{N}&\approx \nbb(\bar x)\,{\mathbf{1}}
+G(x)\Big[ \Theta_I \,{\mathbf{1}}+ \Theta_+\,{\boldsymbol M}_+ + \Theta_-\,{\boldsymbol M}_-\Big]
\nonumber\\[1mm]
& + G(x)\Big[ (\Theta_I^2+\Theta_+\Theta_-) \,{\mathbf{1}}+ 2\Theta_I \Theta_+\,{\boldsymbol M}_+ + 2\Theta_I \Theta_-\,{\boldsymbol M}_-\Big]
\nonumber\\[0mm]
& \;+ Y_{\rm SZ}(x)\Big[ \frac{1}{2}(\Theta_I^2+ \Theta_+\Theta_-) \,{\mathbf{1}}+ \Theta_I \Theta_+\,{\boldsymbol M}_+ + \Theta_I \Theta_-\,{\boldsymbol M}_-\Big].
\label{eq:sup_lin_pol_new_great}
\end{align}
The first term just gives the background blackbody spectrum at the average temperature $\bar T$, while the second term $\propto G(x)$ captures the usual first-order temperature perturbations. The other terms describe second-order corrections with distortion due to superposition of blackbodies at different temperatures with the effect of partial linear polarization included.

From Eq.~\eqref{eq:sup_lin_pol_new_great}, we can obtain the averaged photon occupation number, summed over the polarization states as
\beal
\left<n\right> = \frac{1}{2}\left<{\rm Tr}\mathcal{N} \right>&\approx \nbb(\bar x) 
+ 2 y \,G(x)  + y\, Y_{\rm SZ}(x).
\end{align}
This represents a distorted blackbody at temperature $T'=\bar T (1+2 y)$ [first two terms] with additional $y$-distortion $\propto y\,Y_{\rm SZ}(x)$, where the effective $y$-parameter is 
\beal
\label{eq:y_sup_def}
y =\frac{1}{2}\Big<\Theta_I^2+\Theta_+ \Theta_-\Big> \equiv  \frac{1}{2}\Big<\Theta_I^2+\Theta_Q^2+\Theta_U^2\Big>.
\end{align}
This expression shows that in addition to the temperature perturbations of Stokes $I$, the $y$-parameter also depends on those of $Q$ and $U$. This changes the effective heating rate due to the dissipation of acoustic modes, an effect that was previously neglected. However, these terms only become noticeable at late times, when the tight coupling approximation breaks down (see Sect.~\ref{sec:heating_rate}).

\vspace{-4mm}
\section{Effective heating rate due to damping of scalar perturbations}
\label{sec:heating_rate}
To obtain expressions for the effective heating rate caused by the dissipation of different perturbations, we start by recapping the arguments for scalar perturbations. The physics of the problem is related to the superposition of blackbodies of varying temperatures, where the mixing process is mediated by Thomson scattering. By smoothing fluctuations, Thomson scattering causes an increase of the average CMB temperature by $\Delta T/T\simeq 2 y$ but also sources a $y$-distortion, which subsequently evolves towards a $\mu$-distortion by Compton scattering. The energy injected momentarily as a distortion is $\Delta \rho_\gamma/\rho_\gamma\simeq 4 y$, which corresponds to $1/3$ of the total energy that is converted from the spatially varying part (the acoustic wave) to the smooth average photon field \citep{Chluba2012}. 

The effective heating rate is basically given by the time derivative of the effective $y$-parameter caused by the superposition of blackbodies at different temperatures\footnote{Here, we used the approximation $a^{-4} \rho_\gamma^{-1}\id (a^4 Q)/\id t\approx \id (Q/\rho_\gamma)/\id t$, which is valid since the distortion correction to the photon energy density is very small and $\rho_\gamma\propto a^{-4}$ to high precision. Also, $Q$ is the energy density that is liberated and not to be confused with Stokes $Q$.}:
\beal
\label{eq:heat_definition}
&\left.\frac{\id (Q/\rho_\gamma)}{\id t} \right.\,
\approx - 4\frac{\id}{\id t} y 
=\left.\frac{\!\id (Q/\rho_\gamma)}{\id t}\right|_{\rm I}  + \left.\frac{\!\id (Q/\rho_\gamma)}{\id t}\right|_{\rm P}
\nonumber\\
&\left.\frac{\!\id (Q/\rho_\gamma)}{\id t}\right|_{\rm I} =  - 4 \Big<\Theta_I \dot \Theta_I\Big>
\nonumber\\
&\left.\frac{\!\id (Q/\rho_\gamma)}{\id t}\right|_{\rm P}
=  - 4 \Big<\Theta_Q \dot\Theta_Q+\Theta_U\dot \Theta_U\Big>
= - 2  \Big<\Theta_+ \dot\Theta_- +\Theta_-\dot \Theta_+\Big>,
\end{align}
where for the time derivative only changes due to scattering terms have to be included, i.e. $\dot \Theta_i\rightarrow\dot \Theta_i|_{\rm sc}$ [for scalar perturbations this was shown by \citet{Chluba2012}, but we generalize to vector and tensor perturbations in Appendix~\ref{app:proof}]. This expression neglects corrections due to second-order scattering terms, which ensure that the final heating rate is frame independent. Adding these terms and including only the intensity part of this heating rate, we find \citep{Chluba2012, Chluba2013iso}
\begin{align}
\label{eq:Sac_full_I}
&\!\!\!\!\!\!\!\! \left.\frac{\!\id (Q/\rho_\gamma)}{\id t}\right|_{\rm I} 
=
 4 \taudot\! \int \frac{k^2\!\id
  k}{2\pi^2}P_i(k)
\left[\frac{\left(3\Theta_1-\varv\right)^2}{3}+\frac{9}{2}\Theta_2^2
  \nonumber\right.
  \\
  &\qquad\qquad
  \left.
  -\frac{1}{2}\Theta_2\left(\Theta_0^{\rm P}+\Theta_2^{\rm P}\right)+\sum_{\ell\ge 3}(2\ell+1)\Theta_{\ell}^2\right],
\end{align}
where $\Theta_\ell$ and $\Theta^{\rm P}_\ell$ denote the photon temperature and polarization transfer functions and $\varv$ the one for the baryon velocity \citep[e.g.,][]{Ma1995}. For adiabatic modes, we set the initial power spectrum to $P_i(k)=P_\zeta(k)$, where $P_\zeta(k)$ denotes the power spectrum of curvature perturbations. We furthermore used the time derivative of the Thomson optical depth $\taudot=\sigT \Ne c\approx \pot{4.4}{-21}(1+z)^{3}\,{\rm sec^{-1}}$.

In the derivation of \cite{Chluba2012}, the correction due to the last two terms in Eq.~\eqref{eq:y_sup_def} was not included. These only become noticeable at late times, when the tight coupling approximation breaks down (see Sect.~\ref{sec:TC_normal}); however, for completeness we give them explicitly. Starting from Eq.~(64) of \cite{Ma1995} for the polarization contributions (only $m=0$), we can readily find the additional heating terms
\begin{align}
\label{eq:Sac_full_P}
&\!\!\!\!\left.\frac{\!\id (Q/\rho_\gamma)}{\id t}\right|_{\rm P} 
=
 8 \taudot\! \int \frac{k^2\!\id
  k}{2\pi^2}P_i(k)
  \left[3\left(\Theta_1^{\rm P}\right)^2+\frac{9}{2}\left(\Theta^{\rm P}_{2}\right)^2-\frac{1}{2}\Theta_2\left(\Theta_0^{\rm P}+\Theta_2^{\rm P}\right)
  \nonumber\right.
  \\
  &\qquad\qquad\qquad\quad
  \left.
 +\frac{1}{2}\Theta_0^{\rm P}\left(\Theta_0^{\rm P}-2\Theta_2^{\rm P}\right) +\sum_{\ell\ge 3}(2\ell+1)\left(\Theta^{\rm P}_{\ell}\right)^2\right].
\end{align}
This expression includes the contributions from $Q$ and $U$ (extra factor of 2), which for scalar perturbations only involves $E$-mode patterns at first order in perturbation theory. No additional scattering correction $\propto \varv$ arises, since aberration and Doppler boosting terms \citep[e.g.,][]{Dai2014} lead to higher order corrections.

\vspace{-3.5mm}
\subsection{Tight coupling approximation for adiabatic modes}
\label{sec:TC_normal}
In the tight coupling era ($z\gtrsim 10^4$), one finds \citep{Hu1996anasmall} $\Theta^{\rm P}_{0}\approx (5/4)\Theta_2$, $\Theta^{\rm P}_{2}\approx (1/4) \Theta_{2}$ and zero otherwise, so that the new terms cancel identically, $[\!\id (Q/\rho_\gamma)/\!\id t]|_{\rm P}\approx 0$. Using the tight coupling approximation for $\Theta_{2}$, the approximation for the scalar contribution to the heating rate in the tight coupling limit therefore is \citep[e.g.,][]{Chluba2012, Chluba2013iso}
\beal
\label{Qdot_Sc}
\left. \frac{\id (Q/\rho_\gamma)}{\id t}\right|_{\rm S}
&\approx
\frac{16 c^2}{45 a^2 \taudot} D^2 \!\int \!\frac{k^4 \!\id k}{2\pi^2} P_{\zeta}(k)\, 2 \sin^2(k \rs) \, \expf{-2k^2/\kD^2}.
\nonumber\\[2mm]
&=-2 D^2 \!\int \!\frac{k^2 \!\id k}{2\pi^2} P_{\zeta}(k)\, \sin^2(k \rs) \, \frac{\id}{\id t} \expf{-2k^2/\kD^2}.
\end{align}
for adiabatic modes. Here, we have the mode-specific efficiency factor\footnote{For isocurvature modes, see \citet{Chluba2013iso}.}, $D^2=[1+(4/15)R_\nu]^{-2}$, where $R_\nu=\rho_\nu/(\rho_\gamma+\rho_\nu)\approx 0.41$ is the fractional contribution of massless neutrinos to the energy density of relativistic species. Furthermore, $a=1/(1+z)$ denotes the scale factor normalized to unity today, $\rs=\int \cs \id t/a$ the sound horizon with sound speed $\cs\simeq c /\sqrt{3}$ and $\kD\simeq \pot{4.0}{-6}(1+z)^{3/2} \Mpc^{-1}$ the damping scale, which is determined by $\partial_t \kD^{-2}=8 c^2/[45 a^2 \taudot]$ (neglecting baryon loading). 

Equation~\eqref{Qdot_Sc} is sufficient at $z\gtrsim 10^4$ but becomes inaccurate at later time \citep{Chluba2012}. In particular, around recombination, when polarization anisotropies start to arise, one expects percent level corrections to the heating rate from the new terms, Eq.~\eqref{eq:Sac_full_P}. However, since this only gives rise to a $y$-distortion that is much smaller than the one produced by the formation of structures and reionization era \citep[e.g.,][]{Hu1994pert, Refregier2000}, we do not consider this in more detail here.

For smooth power spectra, $2 \sin^2(k \rs)\approx 1$, which is very accurate for nearly scale invariant perturbations. This expression can be used to compute the spectral distortion for given $P_\zeta(k)$, and limits have been discussed extensively \citep[e.g.,][]{Chluba2012, Chluba2012inflaton, Powell2012, Khatri2013forecast, Chluba2013fore, Chluba2013PCA, Clesse2014}.

\vspace{-0mm}
\section{Heating rate for vectors and tensors}
\label{sec:heating_general}
For vector and tensor perturbations, we can proceed in a similar way as for scalars. According to Eq.~\eqref{eq:heat_definition}, we need to compute the averages $\big<\Theta_I \dot \Theta_I\big>$ and $\!\id \big<\Theta_+ \Theta_-\big>/\!\id t=\big<\Theta_+ \dot \Theta_-\big>+\big<\Theta_- \dot \Theta_+\big>$ using the photon transfer functions in Fourier space. Explicitly, we use the mode decomposition \citep[see][]{Hu1997}
\beal
\label{eq:Fourier_conventions}
\Theta_I(t, \vek{x}, \vek{n})&=\!\!\int \!\!\frac{\id^3 k}{(2\pi)^3} \expf{i \scriptsize \vek{x}\cdot\vek{k}}
\!\sum_\ell \!\sum_{m=-2}^{m=2}  
\!(-i)^\ell \!\sqrt{\frac{4\pi}{2\ell+1}}\,\, {}_0 Y_{\ell m}(\vek{n}) \, \Theta_\ell^{(m)}(\vek{k})
\\
\nonumber
\Theta_{\pm}(t, \vek{x}, \vek{n})&=\!\!\int \!\!\frac{\id^3 k}{(2\pi)^3} \expf{i \scriptsize \vek{x}\cdot\vek{k}}
\!\sum_\ell \!\sum_{m=-2}^{m=2}  
\!(-i)^\ell \!\sqrt{\frac{4\pi}{2\ell+1}} {}_{\pm2}Y_{\ell m}(\vek{n}) \, \Theta^{(m)}_{\pm, \ell}(\vek{k})
\end{align}
with spin harmonics ${}_{s}Y_{\ell m}(\vek{n})$ and $\Theta^{(m)}_{\pm, \ell}(\vek{k})=E_\ell^{(m)}(\vek{k})\pm i B_\ell^{(m)}(\vek{k})$. The averages over photon directions and $\vek{x}$ are explicitly carried out in Appendix~\ref{app:integrals_long}. For statistically isotropic perturbations, with the scattering terms of Eq.~(60), (63) and (64) from \citet{Hu1997} and Eq.~\eqref{eq:TI2_term} and \eqref{eq:Tpm2_term}, after correcting the gauge dependence (i.e., using $\Theta^{(m)}_1\rightarrow \Theta^{(m)}_1-\varv^{(m)}$) we have
\bsub
\label{eq:heating_general}
\beal
\label{eq:heating_general_a}
\left<\Theta^{(m)}_I \dot \Theta^{(m)}_I\right>&\approx - \!\int \!\frac{k^2\!\id k}{2\pi^2} P_{i}^{(m)}(k)\, \taudot \left[
\frac{(\Theta_1^{(m)}-\varv^{(m)})^2}{3}(1-\delta_{m2})\right.
\\
&\qquad \left.
+\frac{\Theta_2^{(m)}}{5}\left(\frac{9}{10}\Theta_2^{(m)}+\frac{\sqrt{6}}{10} E_2^{(m)}\right)
+\sum_{\ell \geq 3} \frac{(\Theta_\ell^{(m)})^2}{(2\ell+1)}
\right]
\nonumber
\\[2mm]
\label{eq:heating_general_b}
\frac{\id\left<\Theta^{(m)}_+ \Theta^{(m)}_-\right>}{\id t}&\approx 
- 2\!\int \!\frac{k^2\!\id k}{2\pi^2} P_{i}^{(m)}(k)\, \taudot 
\left[
\frac{2E_2^{(m)}}{25}\left(\frac{\sqrt{6}}{4} \Theta_2^{(m)}  + E_2^{(m)}\right)
\right.
\nonumber\\
&\quad \qquad\left.
+\frac{1}{5}(B_2^{(m)})^2
+\sum_{\ell\geq 3} \frac{(E_\ell^{(m)})^2+(B_\ell^{(m)})^2}{(2\ell+1)}
\right],
\end{align}
\esub
for each $m$. Here, $P_i^{(m)}$ denote the initial power spectra for scalar ($m=0$), vector ($m=\pm1$) and tensor ($m=\pm2$) perturbations. Then, the total heating rate directly follows from Eq.~\eqref{eq:heat_definition} after summing over $m$. Assuming that $P_i^{(m)}=P_i^{(-m)}$, this produces another factor of $2$ for $m\neq 0$. The correspondence of Eq.~\eqref{eq:heating_general_a} with Eq.~\eqref{eq:Sac_full_I} can be shown using $\Theta^{(0)}_\ell = (2\ell +1)\, \Theta_\ell$, $\Theta^{\rm P}_\ell = 4 G^{(0)}_\ell$ and $E_2^{(0)}=-\frac{5}{\sqrt{6}}[\Theta^{\rm P}_0+\Theta^{\rm P}_2]$ \citep[see,][]{Tram2013}. Similarly, the correspondence of Eq.~\eqref{eq:heating_general_b} and Eq.~\eqref{eq:Sac_full_P} can be shown with Eq.~(B.11) of \citet{Tram2013} and $B^{(0)}_\ell=0$, even if the derivation is lengthy.

Since vector perturbations are usually not excited by inflation, we defer a more detailed discussion to another paper and now restrict our attention to tensor contributions. However, Eq.~\eqref{eq:heating_general} provides the general expression including the effect of all polarization states and higher multipoles. It thus can be used to describe the heating rate for general perturbations sourced by scalar, vector and tensor perturbations, once the transfer function $\Theta_\ell^{(m)}$, $\varv^{(m)}$, $E_\ell^{(m)}$ and $B_\ell^{(m)}$ are available. Below, we only discuss approximate solutions for the tensor transfer function, but the results emphasize the most relevant physical aspects.

\vspace{-2mm}
\subsection{Tight coupling approximation for tensor perturbations}
\label{sec:Tensor_TC_I}
For scalar perturbations, a precise approximation for the effective heating rate can be obtained in the tight coupling limit (see Sect.~\ref{sec:TC_normal}). We will see that for tensors, corrections caused in the quasi-free streaming regime\footnote{`Quasi' because for full free streaming there is no mixing by scattering.} at very small scales need to be included (Sect.~\ref{sec:quasi-tight}), but for a basic understanding even the tight coupling approximation is sufficient.

The tight coupling solutions for scalar, vector and tensor perturbations are discussed in Sect. IV of \cite{Hu1997} and a brief derivation is given in Appendix~\ref{app:ana_sol_env}. In this limit, $B_\ell^{(m)}\approx 0$ and we need not worry about their contribution any further. Also, there is no $\ell=1$ contribution for $m=2$. For $E_\ell^{(m)}$, the only non-vanishing component is, $E_2^{(m)}\approx -\sqrt{6}\Theta_2^{(m)}/4$, and thus, $\!\id\big<\Theta^{(m)}_+ \Theta^{(m)}_-\big>/\!\id t\approx 0$ [cf. Eq.~\eqref{eq:heating_general_b}]. With $\frac{9}{10}\Theta_2^{(m)}+\frac{\sqrt{6}}{10} E_2^{(m)}\approx (3/4) \Theta_2^{(m)}$, from Eq.~\eqref{eq:heating_general_a}, the only non-vanishing term is
\beal
\label{eq:Qdot_T_in}
\left.\left<\Theta_I \dot \Theta_I\right>\right|_{\rm T}
&\approx - \int \!\frac{k^2\!\id k}{2\pi^2} P_{i}^{(2)}(k)\, \frac{3\taudot}{10}
 \left(\Theta_2^{(2)}\right)^2
 \nonumber\\[2mm]
&\approx- \int \!\frac{k^2\!\id k}{2\pi^2} P_{h}^{(2)}(k)\, \frac{8\dot h^2}{15\taudot},
\end{align}
where we multiplied by $2$ assuming $P_{i}^{(2)}(k)=P_{i}^{(-2)}(k)$. In the second line, we used the tight coupling approximation, $\Theta_2^{(2)}\approx -\frac{4}{3}\frac{\dot h}{\taudot}$ [Eq.~(91) of \citet{Hu1997}], where $h$ parametrizes the amplitude of the tensor mode, and identified $P_{i}^{(2)}(k)=P_{h}^{(2)}(k)$. Now the goal is to relate the power spectrum of $h$ with the tensor power spectrum, $P_T(k)$, defined for both helicities. From the definitions of \citet{Hu1997}, it follows that for one helicities of $h$ we have $P^{(2)}_{h}(k)=[(3/8)\times 4]^{-1} \times (1/4) \times [P_T(k)/2]=P_T(k)/12$, where the first factor arises due to the normalization of the eigenfunctions $Q^{(2)}_{ij}$ and the factor of $(1/4)$ from the factor of $2$ in $h_{ij}=2 h Q^{(2)}_{ij}$. Finally, $P_T(k)/2$ simply gives the power for one helicity.

\begin{figure}
\centering
\includegraphics[width=1.02\columnwidth]{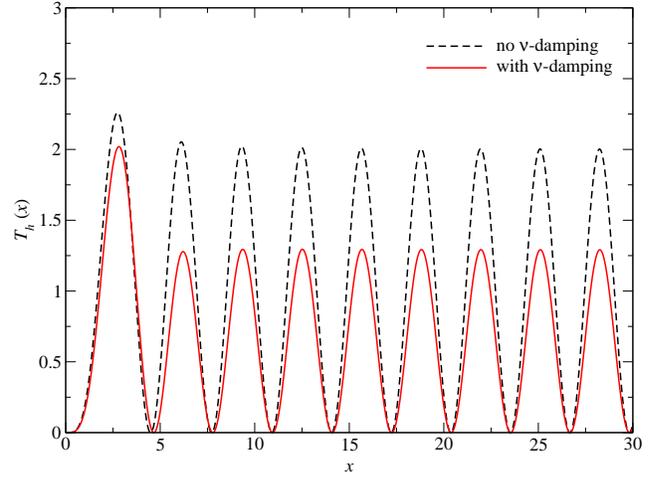}
\caption{Tensor transfer function $\mathcal{T}_h(x)$. The dashed line shows the case without neutrino damping, while for the solid line it was included.}
\label{fig:Tfunc}
\end{figure}

To obtain an expression for the tensor contribution to the heating rate, we need approximations for the transfer function of $h(t, k)$. In this section, we only consider energy release before the $y$-era ($z\gtrsim10^4$), which is still radiation dominated. Energy release at $z\lesssim 10^4$ is discussed in Sect.~\ref{sec:late_release}.
In this regime, the damping caused by the free streaming of neutrinos has to be included \citep{Weinberg2004}, which suppresses the amplitude of $h$ by $\simeq 20\%$ for $k\eta\gg1$ \citep[for the tensor power, this makes a difference of $\simeq 36\%$;][]{Weinberg2004}. Neutrinos only start free streaming after decoupling at a temperature $T\simeq 1.5\,{\rm MeV} -2\,{\rm MeV}$, or redshift $z\simeq \pot{6}{9}-\pot{9}{9}$. Before that, they are coupled to the plasma and their contribution to the anisotropic stress remains small. Free streaming of neutrinos at $z\gtrsim 10^4$ thus mostly affects scales $\pot{2}{-2}\,\Mpc^{-1}\lesssim k \lesssim \pot{\rm few}{4}-10^5\,\Mpc^{-1}$ \citep[e.g.,][]{Boyle2008b, Watanabe2006}.

The tensor power spectrum furthermore is modified by changes of the effective number of relativistic degrees of freedom during the electron-positron annihilation and the quark-gluon phase transition \citep{Watanabe2006}. This introduces several features into the tensor power spectrum at small scales \citep[see Figs.~4 and 5 of][]{Watanabe2006}, however, we neglect these complications, which are only noticeable (at the level of $\simeq 20\%-30\%$) for very blue tensor power spectra, and just include the effect of neutrino free streaming at all small scales. With this simplification, we find that at $200\,\Mpc^{-1}\lesssim k \lesssim \pot{2}{4}\,\Mpc^{-1}$, the tensor power is on average overestimated by $\simeq 10\%-20\%$. At $\pot{2}{4}\,\Mpc^{-1}\lesssim k \lesssim 10^6\,\Mpc^{-1}$, the power is underestimated by a factor of $\simeq 1.5$, while at $10^6\,\Mpc^{-1}\lesssim k \lesssim 10^9\,\Mpc^{-1}$, it is overestimated $\simeq 1.5$ times \citep[cf. Fig.~5 of][]{Watanabe2006}.

A simple analytic expression for $h(t, k)$ that include the effect of neutrino damping was derived by \citet{Dicus2005}. Including the small correction to $\dot h= h' c / a$ due to photon damping\footnote{Although energetically this does not make a significant difference, the extra factor of $\expf{-\Gamma_\gamma \eta}$ is the origin of the heating, as we explain below. It also emphasizes the similarities to the heating rate for adiabatic modes, Eq.~\eqref{Qdot_Sc}.} (see Appendix~\ref{app:gamma_damp}), with Eq.~\eqref{eq:sol_h_damping} and \eqref{eq:Qdot_T_in} we can approximate the tensor contribution to the heating rate as (see Sect.~\ref{sec:consistency} for an alternative derivation)
\beal
\label{eq:Qdot_T}
\left.\frac{\id (Q/\rho_\gamma)}{\id t}\right|_{\rm T}
&\approx
\frac{4 H^2}{45 \taudot} \int_0^{k_{\rm cut}} 
\frac{k^2 \!\id k}{2\pi^2} P_T(k)\,  \mathcal{T}_h(k\eta) \,\expf{-\Gamma_\gamma \eta}
\nonumber\\
&= - \frac{1}{24(1-R_\nu)}\,\int_0^{k_{\rm cut}} 
\frac{k^2 \!\id k}{2\pi^2} P_T(k)\,  \mathcal{T}_h(k\eta) \,\frac{\id}{\id t}\,\expf{-\Gamma_\gamma \eta}
\nonumber\\[1mm]
\mathcal{T}_h(x)&\approx 2\left\{\sum^6_n a_n [n j_{n}(x)-x j_{n+1}(x)] \right\}^2,
\end{align}
where $j_n(x)$ denote spherical Bessel functions with the numerical coefficients $a_0=1$, $a_2=0.243807$, $a_4=\pot{5.28424}{-2}$ and $a_6=\pot{6.13545}{-3}$ and $\id (\Gamma_\gamma \eta)/\id t=32 H^2 (1-R_\nu) /[15 \taudot]$. We also introduced a cutoff scale $k_{\rm cut}$ (to regularize the integral), which we discuss below, and assumed radiation domination so that $H\approx c/(a \eta)$.
The dependence of $\mathcal{T}_h(x)$ on $x$, both with and without the effect of neutrinos, is shown in Fig.~\ref{fig:Tfunc}. The contribution at small scales is overestimated $\simeq 1.5$ times if neutrino damping is neglected. At $k\eta\gtrsim 5$, one has $\mathcal{T}_h(x)\simeq 1.29 \cos^2(k\eta)$.

%
For $P_{\rm T}=2\pi^2 A_{\rm T} k^{-3} (k/k_0)^{n_T}$, the integrand of Eq.~\eqref{eq:Qdot_T} scales as $\simeq k^{n_T} k^3$ as $k\rightarrow 0$, while for $k \eta \gg 1$ we have $\simeq k^{n_T-1} \cos^2(k \eta)$. At large scales, $\dot h$ vanishes, so that no super-horizon heating occurs. However, at small scales, we need to introduce a cutoff scale, $k_{\rm cut}$, to regularize the integral. For $\nT=0$, the dependence on the cutoff scale is only logarithmic, but for $\nT>0$ it becomes rather strong (cf. Sect.~\ref{sec:simp_express_Qdot}). One scale is due to the end of inflation and reheating, $k_{\rm end}\simeq 10^{23}\,\Mpc^{-1}$ \citep[e.g.,][]{Boyle2008a}, however, a much larger scale is related to the photon mean free path, $\lambda_{\rm mfp}/a\simeq (\sigT \Ne a)^{-1}$ or $k_{\rm cut}=\sigT \Ne a \simeq \pot{4.5}{-7}(1+z)^2 \,\Mpc^{-1}$. At smaller scales, photons stream quasi-freely, undergoing very few scatterings and adding little extra heating, as we explain below. At redshifts $z\simeq 10^4-\pot{2}{6}$ (relevant for the non-$y$ distortion), we thus have $k_{\rm cut}\simeq 45\,\Mpc^{-1} - \pot{\rm few}{6}\,\Mpc^{-1}$. In contrast, for scalar perturbations, only modes with wavenumber $k\lesssim \pot{\rm few}{4}\,\Mpc^{-1}$ are important. Spectral distortions hence allow probing tensor perturbations to significantly smaller scales (see Fig.~\ref{fig:window}), simply because for scalar perturbations Silk damping erases all temperature fluctuations before they can even reach the quasi-free streaming phase.

\begin{figure}
\centering
\includegraphics[width=1.03\columnwidth]{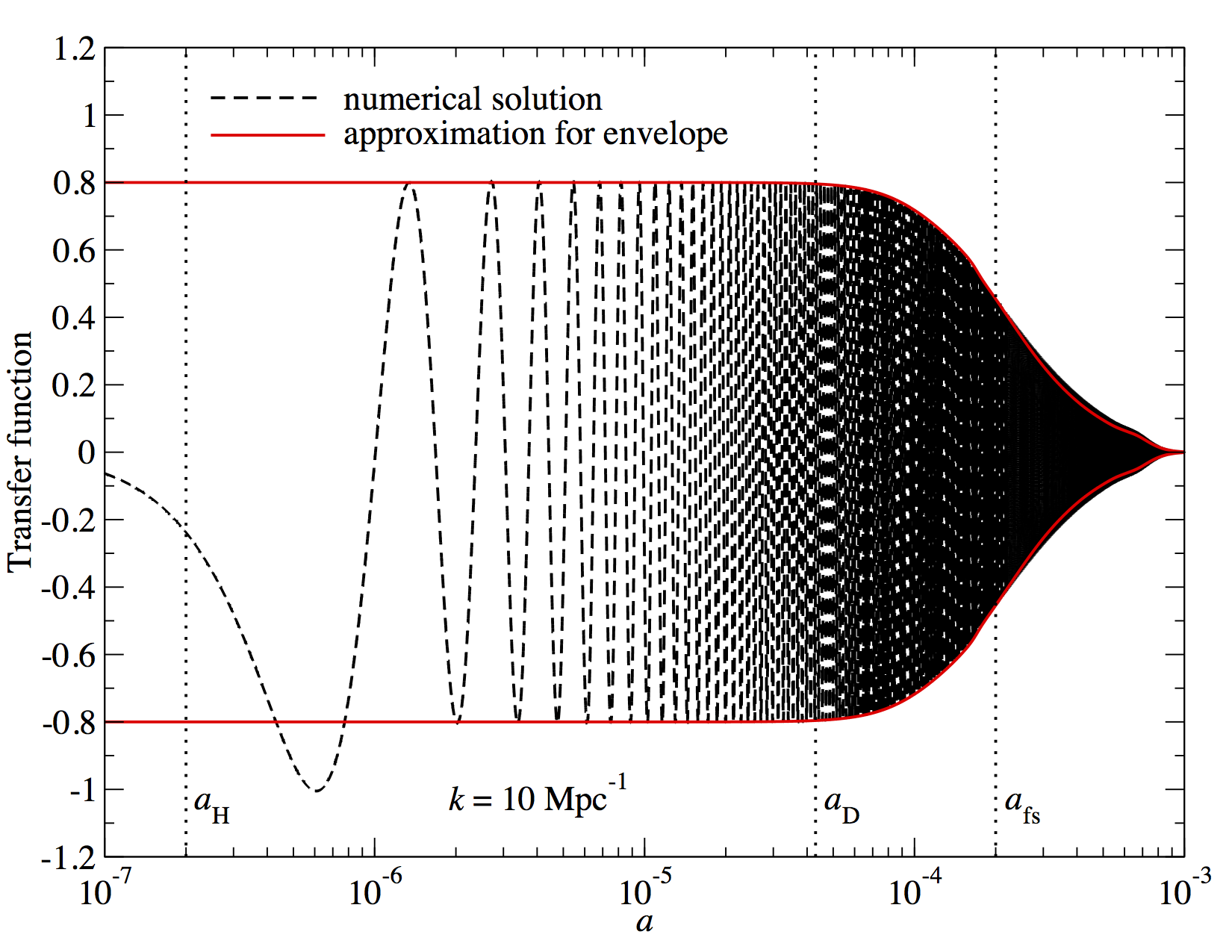}
\\[1.5mm]
\includegraphics[width=1.03\columnwidth]{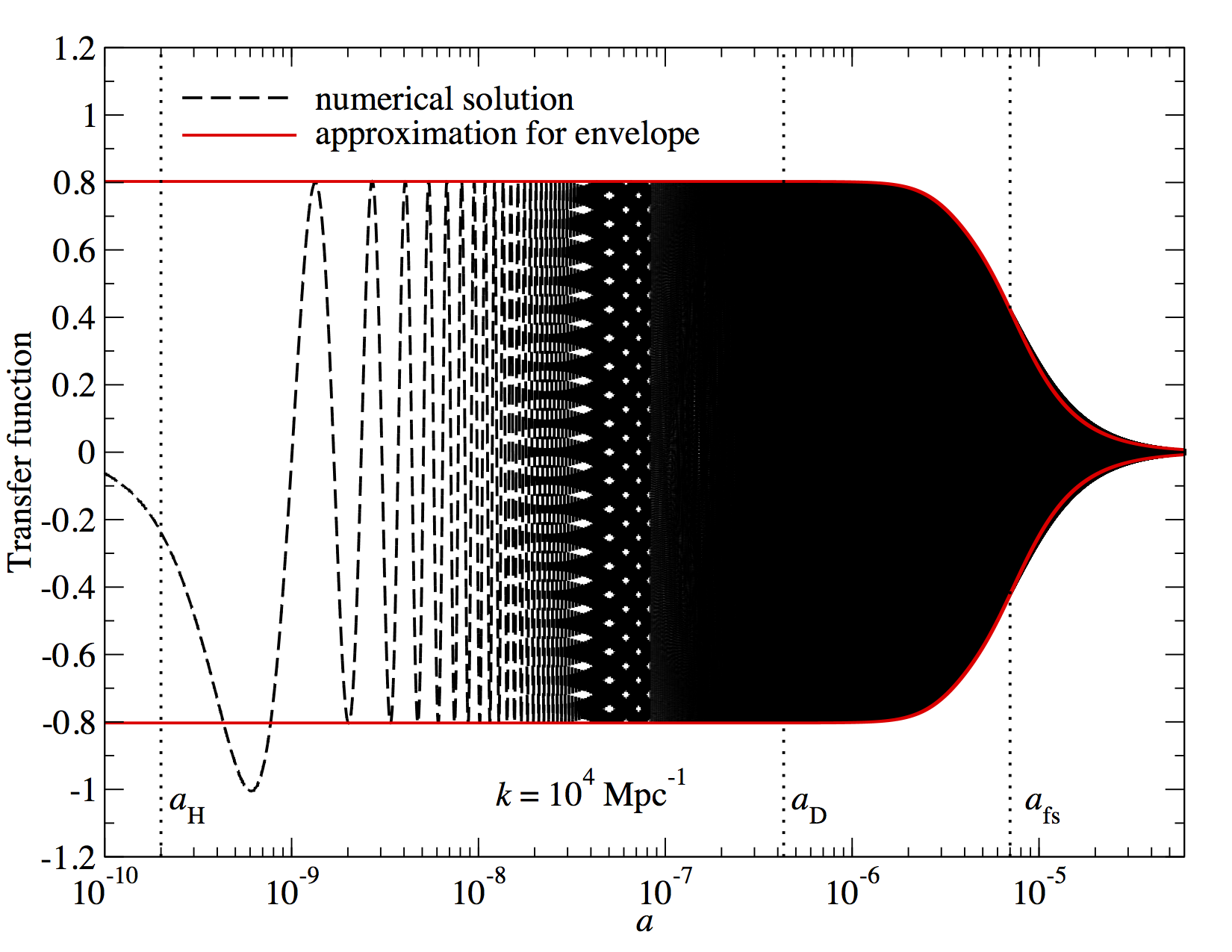}
\caption{Transfer function $\mathcal{T}^{(2)}_2$ at $k=10\,\Mpc^{-1}$ and $k=10^4\,\Mpc^{-1}$. For $k\ll \tau'$, the tight coupling approximation describes the solution very well, while later the response of the photon field becomes weaker. The envelope of the solution can be represented with the approximation Eq.~\eqref{eq:appro_transfer_functions_a}, multiplied by $\sqrt{1.29/2}\simeq 0.8$ to account for the suppression of the tensor amplitude by neutrino damping after horizon crossing. The vertical lines indicate the scale factor at horizon crossing, $a_{\rm H}$, and when the mode reaches the diffusion scale for scalar modes, $a_{\rm D}$, and free streaming scale, $a_{\rm fs}$.}
\label{fig:TE_funcs}
\end{figure}

\vspace{0mm}
\subsection{Quasi-tight coupling approximation for tensors}
\label{sec:transferfunctions_tensors}
\label{sec:quasi-tight}
To improve the approximation for the effective heating rate caused by tensor perturbations, we need to include radiative transfer effects at small scales, when photons approach the free streaming regime. The evolution of fluctuations in the photon field that are sourced by tensor perturbations is generally simpler than for scalars. Tensors only excite modes with $m=\pm 2$ and in contrast to scalar perturbations, the effect of photons on the amplitude of the tensor perturbations is negligible. Thus, the photon transfer functions are characterized by the driving force of the tensor fluctuations at all phases of the evolution, while for scalar perturbations, the potentials quickly disappear after entering the horizon. 

With the analytic solution, Eq.~\eqref{eq:sol_h_damping}, for the tensor amplitude $h$ in the radiation-dominated era, we can numerically solve the photon Boltzmann hierarchy given in \citep{Hu1997} for $\Theta^{(2)}_\ell$, $E^{(2)}_\ell$ and $B^{(2)}_\ell$. We modified the Boltzmann solver of {\sc CosmoTherm} \citep{Chluba2011therm} for this purpose.
For $k\ll \tau'$, where $\tau'=(a/c) \taudot$,  we are in the tight coupling regime having $\Theta^{(2)}_2\simeq -(4/3) \,h'/\tau'$ and $E^{(2)}_2\simeq - \sqrt{6} \Theta^{(2)}_2/4=\sqrt{2/3}\, h'/\tau'$. To discuss numerical solutions and the corrections caused by radiative transfer effects, it is thus useful to introduce the transfer functions 
\beal
\nonumber
\mathcal{T}^{(2)}_\ell&=\frac{\Theta^{(2)}_\ell}{-4/(3\eta\tau')}, 
&\mathcal{E}^{(2)}_\ell&=\frac{E^{(2)}_\ell}{-4/(3\eta\tau')},
&\mathcal{H}^{(2)}_\ell&=\frac{B^{(2)}_\ell}{-4/(3\eta\tau')}.
%
\end{align}
Here, we set the initial amplitude of $h$ to unity and used the dominant scaling with conformal time, $h'\simeq A_h/\eta$.

In Fig.~\ref{fig:TE_funcs}, we illustrate the transfer functions for $\Theta^{(2)}_2$ at wavenumber $k=10\,\Mpc^{-1}$ and $k=10^4\,\Mpc^{-1}$. We included photon perturbations up to $\ell=10$ and assumed a standard cosmology \citep{Planck2013params} for the numerical computation. We computed the recombination history with {\sc CosmoRec} \citep{Chluba2010b}. For $k\ll \tau'\propto \eta^{-2}$, the tight coupling approximation describes the solution very well, while later the response of the photon field becomes much weaker. In this regime, photons stream quasi-freely and the response to the driving force becomes weaker even if tensor modes are still present and wiggling around, attempting to excite temperature and polarization anisotropies. The problem becomes similar to a system of driven damped oscillators that become more weakly coupled. The transition from tightly coupled to weakly coupled occurs around $a_{\rm fs}\simeq \pot{7}{-4} (\Mpc^{-1}/ k)^{1/2}$, which for $k=10\,\Mpc^{-1}$ is $a_{\rm fs}\simeq \pot{2}{-4}$ and $a_{\rm fs}\simeq \pot{7}{-6}$ for $k=10^4\,\Mpc^{-1}$. In contrast, for the diffusion scale of scalar modes, we have $a_{\rm D}\approx \pot{2}{-4} (k\,\Mpc)^{-2/3}$, implying $a_{\rm D}\approx \pot{4.3}{-5}$ and $a_{\rm D}\approx \pot{4.3}{-7}$, respectively.

With this picture in mind, one can find simple approximations for the envelope of the transfer functions, as explained in Appendix~\ref{app:ana_sol_env}. These approximations clearly capture the solution for $\Theta^{(2)}_2$ very well (see Fig.~\ref{fig:TE_funcs}), even close to the recombination era. 
In the quasi-free streaming phase, the approximation slightly underestimates the envelope of the numerical solution. This is because we only included multipoles $\ell=2$, but better agreement can be achieved by adding the term for $\ell=3$ (Appendix~\ref{app:high_prec}).
We also find the approximations for $E^{(2)}_2$ and $B^{(2)}_2$ to reproduce our numerical results very well, but their contribution to the heating is generally smaller. The amplitude of $\Theta^{(2)}_2$ decays as $\simeq \tau'/k$, while the one for $E^{(2)}_2$ declines faster $\simeq (\tau'/k)^2$. This decay is much slower than for scalar perturbations, which damp exponentially $\simeq \exp(-k^2/\kD^2)$ by photon diffusion. 
In the free streaming regime, also modes with $\ell>2$ are excited, but overall these add a smaller correction [a few percent for nearly scale invariant tensor power spectrum (Sect.~\ref{sec:mu_res})] to the heating rate and thus can usually be neglected. In Sect.~\ref{sec:higher_corrs}, we shall include these corrections quasi-analytically. 

To obtain the solutions for the photon transfer functions, we introduced a hard cut at $\ell_{\rm max}$, setting multipoles with $\ell>\ell_{\rm max}$ to zero. We find that the transfer functions coverage very rapidly at all phases of the evolution relevant to us  when changing $\ell_{\rm max}$. For example, $\mathcal{T}^{(2)}_2$ changes only minimally when going from $\ell_{\rm max}=2$ to $3$, and changing to $\ell_{\rm max}=10, 20$ and $40$, already makes practically no difference. The photon fluid simply does not support shear waves at first order in perturbation theory, so that the error introduced by truncating the mode hierarchy does not propagate very strongly. We also find that the amplitude of the transfer functions for higher multipoles drops rapidly in the free streaming regime. This means that higher multipoles only add a tiny amount of extra heating, implying that also the heating rate converges very rapidly with $\ell_{\rm max}$ (cf. Fig.~\ref{fig:heating_one_mode_comp} and Sect.~\ref{sec:higher_corrs}).

\vspace{0mm}
\subsubsection{Improved tight coupling approximation}
With this more detailed understanding of the photon transfer effects at small scales, we can improve the approximation for the heating rate. In particular, we do not need to add any cutoff scale by hand, since free streaming corrections naturally limit the contributions to the heating from small scales. The more accurate heating rate reads
\beal
\label{eq:Qdot_T_improved}
\left.\frac{\id (Q/\rho_\gamma)}{\id t}\right|_{\rm T}
&\approx
\frac{4H^2}{45 \taudot} \int_0^\infty
\frac{k^2 \!\id k}{2\pi^2} P_T(k)\,  \mathcal{T}_h(k\eta) \,\mathcal{T}_\Theta(k/\tau')\,
\expf{-\Gamma^\ast_\gamma(k, \eta)\,\eta}
\nonumber\\
&= - \frac{1}{24 (1-R_\nu)}\,\int_0^{k_{\rm cut}} 
\frac{k^2 \!\id k}{2\pi^2} P_T(k)\,  \mathcal{T}_h(k\eta) \,\frac{\id}{\id t}\,\expf{-\Gamma^\ast_\gamma(k, \eta)\,\eta}
\nonumber\\[2mm]
\mathcal{T}_\Theta(\xi)&= 
\frac{1+\frac{341}{36}\xi^2 + \frac{625}{324}\xi^4}{1+\frac{142}{9}\xi^2 + \frac{1649}{81}\xi^4 + \frac{2500}{729} \xi^6},
\end{align}
where the scale-dependent damping coefficient is determined by 
\beal
\label{eq:damping_coefficient}
\frac{\id (\Gamma^\ast_\gamma \eta)}{\id t}=\frac{32 H^2 [1-R_\nu]\, \mathcal{T}_\Theta(k/\tau')}{15 \taudot}.
\end{align}
To obtain Eq.~\eqref{eq:Qdot_T_improved}, we only used the transfer function for $\Theta^{(2)}_2$, replacing $\mathcal{T}^{(2)}_2=1$, which was used for the approximation Eq.~\eqref{eq:Qdot_T}, with the more accurate expression from Eq.~\eqref{eq:appro_transfer_functions_a}. We can see that for $k\gg \tau'$, the integrand of Eq.~\eqref{eq:Qdot_T_improved} scales as $k^{\nT-1} \cos^2 (k\eta) [\tau'/k]^2$, so that for $\nT<2$ the integral converges. Due to the oscillatory behavior of $\mathcal{T}_h(k\eta)$, in practice for $k\eta \gg 1$ we use the averaged value, $\big<\mathcal{T}_h(k\eta)\big>\approx 1.29/2$, over one oscillation phase. This eases the numerical evaluation of the heating rate and does not make much of a difference for smooth power spectra.

\begin{figure}
\centering
\includegraphics[width=1.03\columnwidth]{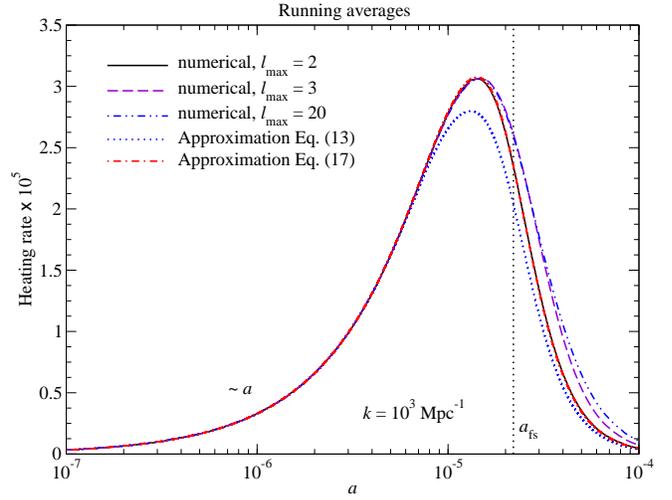}
\caption{Averaged single-mode heating rate $\!\id (Q/\rho_\gamma)/\!\id \ln z$ (amplitude $A_T=12$) computed numerically for $k=10^3\,\Mpc^{-1}$ and different values of $\ell_{\rm max}$. For comparison, we show the result obtained with the approximations Eq.~\eqref{eq:Qdot_T_improved} and Eq.~\eqref{eq:T_Qdot_l2}. We also indicated the location of $a_{\rm fs}$.}
\label{fig:heating_one_mode_comp}
\end{figure}

In Fig.~\ref{fig:heating_one_mode_comp}, we show the single-mode heating rate averaged over one period for $k=10^3\,\Mpc^{-1}$. At early times, the single-mode heating rate scales as $\id (Q/\rho_\gamma)/\id \ln z \simeq a$ in all cases. Including all terms up to $\ell_{\rm max}=2$ for the numerical calculation, we see that Eq.~\eqref{eq:Qdot_T_improved} underestimates the heating rate by some $\simeq 10\%$. This is because at this point we neglected corrections due to $E^{(2)}_\ell\neq - \sqrt{6} \Theta^{(2)}_2/4$ and $B^{(2)}_\ell\neq 0$, which become noticeable in the free streaming regime. These contributions can also be included analytically, as we show next.

\vspace{0mm}
\subsubsection{Adding all terms for $\ell=2$}
To obtain $\mathcal{T}_\Theta(\xi)$ in Eq.~\eqref{eq:Qdot_T_improved}, we set $B^{(2)}_2=0$ and $E^{(2)}_2=- \sqrt{6} \Theta^{(2)}_2/4$, so that only the transfer function, Eq.~\eqref{eq:appro_transfer_functions_a}, for $\Theta^{(2)}_2$ was needed. However, with the expressions Eq.~\eqref{eq:appro_transfer_functions}, we can also add the corrections for $B^{(2)}_2\neq 0$ and $E^{(2)}_2\neq - \sqrt{6} \Theta^{(2)}_2/4$ in the free streaming phase.
For this, we need to account for the phase difference between $\Theta^{(2)}_2$ and $E^{(2)}_2$, Eq.~\eqref{eq:phases_a} and \eqref{eq:phases_b}, which is important for the cross terms $\propto \Theta^{(2)}_2 E^{(2)}_2$ in the heating rate, Eq.~\eqref{eq:heating_general}. Including all terms up to $\ell_{\rm max}=2$, we have the correction,
\beal
\label{eq:Qdot_l2}
\left.\frac{\id (Q/\rho_\gamma)}{\id t}\right|_{\rm T, c}
&\approx 
4\!\int \!\frac{k^2\!\id k}{2\pi^2} P_{i}^{(2)}(k)\, 
\frac{3\taudot}{10}
\Bigg[\frac{1}{5}\left(\Theta_2^{(2)}\right)^2+\frac{8}{15}\left(E_2^{(2)}\right)^2
\nonumber\\
&\qquad\quad +\frac{4}{3}\left(B_2^{(2)}\right)^2+\frac{8\,\Theta_2^{(2)} E_2^{(2)}}{5\sqrt{6}},
\Bigg]
\end{align}
to the heating rate, Eq.~\eqref{eq:Qdot_T_improved}. Assuming rather smooth tensor power spectra, with the approximations Eq.~\eqref{eq:appro_transfer_functions}, we then find
\bsub
\beal
\label{eq:Theta_l2_a}
\frac{1}{5}\left(\Theta_2^{(2)}\right)^2 &+\frac{8}{15}\left(E_2^{(2)}\right)^2+\frac{4}{3}\left(B_2^{(2)}\right)^2 
\approx
\frac{2}{5} \mathcal{T}_h(k\eta) \, \mathcal{T}_{\Theta, c}(k/\tau')
\nonumber\\[1mm]
\mathcal{T}_{\Theta, c}(\xi )&= 
\frac{1+\frac{139}{24}\xi^2 + \frac{625}{648}\xi^4}
{1+\frac{142}{9}\xi^2 + \frac{1649}{81}\xi^4 + \frac{2500}{729} \xi^6}.
\end{align}
For the cross term between $\Theta_2^{(2)}$ and $E_2^{(2)}$, we have
\beal
\frac{8\,\Theta_2^{(2)} E_2^{(2)}}{5\sqrt{6}}&\approx 
- \frac{2}{5} \mathcal{T}_h(k\eta) \, \mathcal{T}_{\Theta E, c}(k/\tau') \,\cos \delta
\nonumber\\[1mm]
\mathcal{T}_{\Theta E, c}(\xi )&=\frac{\sqrt{1+\frac{377}{36}\xi^2 + \frac{1847}{162}\xi^4+ \frac{625}{324}\xi^6}}
{1+\frac{142}{9}\xi^2 + \frac{1649}{81}\xi^4 + \frac{2500}{729} \xi^6}
\nonumber\\
\delta(\xi)&=
\begin{cases}
\phi_\Theta(\xi)-[\phi_E(\xi)-\pi] & \text{for}\, x \leq 0.7162
\\
\phi_\Theta(\xi)-\phi_E(\xi) &\text{for}\, x > 0.7162
\end{cases},
\end{align}
\esub
where $\phi_\Theta$ and $\phi_E$ are determined by Eq.~\eqref{eq:phases_a} and \eqref{eq:phases_b}. 
For the final approximation, we only need to replace 
\beal
\label{eq:T_Qdot_l2}
\mathcal{T}_\Theta(\xi)\rightarrow \mathcal{T}_\Theta(\xi)
+\frac{2}{5}\Big[\mathcal{T}_{\Theta, c}(\xi)-\mathcal{T}_{\Theta E, c}(\xi) \,\cos \delta\Big]
\end{align}
in Eq.~\eqref{eq:Qdot_T_improved}. We can immediately verify, that in the tight coupling regime ($k/\tau'\ll 1$), we have $\mathcal{T}_{\Theta, c}(k/\tau') \approx \mathcal{T}_{\Theta E, c}(k/\tau') \,\cos \delta\approx 1$, so that the correction terms cancel identically. Comparing with the numerical result for $\ell_{\rm max}=2$, we find that this approximation works extremely well (see Fig.~\ref{fig:heating_one_mode_comp}).

\vspace{-0mm}
\subsubsection{Corrections from higher $\ell$ and polarization states}
\label{sec:higher_corrs}
As mentioned above, in the quasi-free streaming regime also terms from $\ell>2$ start becoming important. This is illustrated in Fig.~\ref{fig:heating_one_mode_comp} for $\ell_{\rm max}=3$ and $\ell_{\rm max}=20$. We can clearly see that even the approximation for $\ell_{\rm max}=2$ already captures most of the heating. Adding terms for $\ell=3$ improves the convergence and very little changes when adding terms up to $\ell_{\rm max}=20$. 
We checked the convergence for even higher $\ell_{\rm max}$ but found no significant difference for the single-mode heating rate.
At late times, the corrections from larger $\ell$ become noticeable, changing the heating rate by a factor of $\simeq 2 - 3$ with respect to $\ell_{\rm max}=2$. For the total single-mode heating rate, this only adds a $\simeq 10\%$ correction, so that one expects to obtain very good results already when using the approximations Eq.~\eqref{eq:Qdot_T_improved} and Eq.~\eqref{eq:T_Qdot_l2}.

It is in principle possible to add terms for $\ell\geq 3$ analytically, using our method in Appendix~\ref{app:ana_sol_env}. However, the expressions become rather cumbersome, so that it is simpler to directly approximate the transfer function, which mainly depends on $\xi=k/\tau'$. We find
\beal
\label{eq:quasi-exact}
\mathcal{T}_{\Theta}(\xi)\approx \frac{1+4.48 \xi+91.0\xi^2}{1+4.64\xi+90.2\xi^2+100\xi^3+55.0\xi^4}
\end{align}
to reproduce the numerical result for $\ell_{\rm max}=20$ and $k=10^3\,\Mpc^{-1}$ extremely well. We checked that this approximation also works very well for even smaller scales. This greatly simplifies the computation of the tensor heating rate.
We also computed the tensor heating rate neglecting the new terms in Eq.~\eqref{eq:heating_general_b}, finding them to change the total result only at the percent level for $k\gtrsim \tau'$.

\vspace{-2mm}
\subsection{Results for the heating rate from tensors}
\label{sec:results_heating_rate}
\begin{figure}
\centering
\includegraphics[width=1.02\columnwidth]{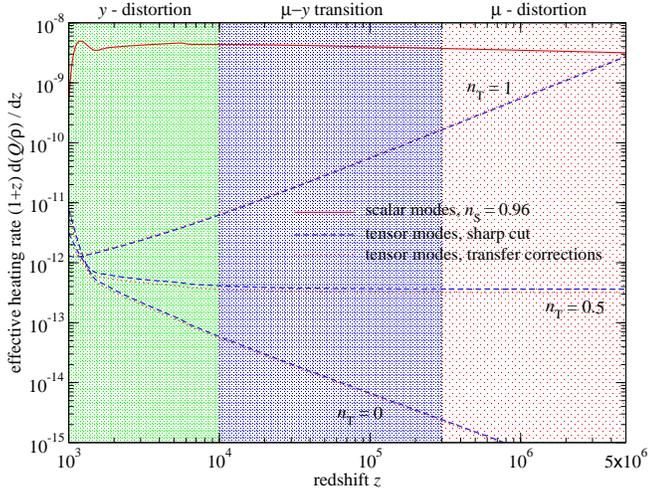}
\caption{Heating rate for tensor modes and different $\nT$. The tensor amplitude was fixed to $A_T=0.1 A_\zeta\approx \pot{2.2}{-10}$ at pivot scale $k_0=0.05\,\Mpc^{-1}$. The blue lines are obtained using $k_{\rm cut}\simeq  \tau'$, in Eq.~\eqref{eq:Qdot_T}, while the red dotted lines include transfer effects using, Eq.~\eqref{eq:Qdot_T_improved}. 
For comparison, we show the heating rate for adiabatic modes using a power spectrum without running. The shaded regions indicate the $y$-era ($z\lesssim10^4$), the $\mu-y$ transition regime ($10^4\lesssim z\lesssim \pot{3}{5}$) and the $\mu$-era ($z\gtrsim \pot{3}{5}$). The tensor heating in the late $y$-era ($z\lesssim \pot{3}{3}$) is overestimated in our computation (see Sect.~\ref{sec:late_release}).}
\label{fig:heat_rate_examples}
\end{figure}

Assuming that the initial tensor perturbations are described by a simple power law, $P_{T}=2\pi^2 A_{\rm T} k^{-3} (k/k_0)^{n_T}$, we can compute the heating rate as a function of redshift. For the tensor amplitude, we use $A_T= 0.1 A_\zeta \approx \pot{2.2}{-10}$ at pivot scale $k_0=0.05\,\Mpc^{-1}$, which is consistent with the upper limit on the tensor to scalar ratio $r\lesssim 0.11$ (95\% c.l.) from Planck \citep{Planck2013params}. 
In Fig.~\ref{fig:heat_rate_examples}, we show the comparison for $\nT=0$, $\nT=0.5$ and $\nT=1$, also varying the approximations used for the transfer functions, Eq.~\eqref{eq:Qdot_T} and Eq.~\eqref{eq:Qdot_T_improved}. The approximations for the heating rate give very similar results, showing that the details of the free streaming corrections are not as important. Using the quasi-exact approximation, Eq.~\eqref{eq:quasi-exact}, we find that for $\nT\lesssim 0.5$, the results for the heating practically coincide with those of Eq.~\eqref{eq:Qdot_T_improved}. For larger $\nT$, the difference can be as large as a factor of $\simeq 1.5$, implying that the heating rate is underestimated by $\simeq 30\%$.
Thus, Eq.~\eqref{eq:Qdot_T}, is sufficient for estimates of the distortion amplitude, while for higher precision Eq.~\eqref{eq:quasi-exact} should be used. The heating at $z\lesssim 10^4$ will be discussed in more detail in Sect.~\ref{sec:late_release}, but is found to be insignificant.

For scale invariant tensor perturbations ($\nT\simeq 0$), the heating rate scales as $\id (Q/\rho_\gamma)/\id \ln z \simeq H/\taudot \simeq 1/(1+z)$. This is because the integrals in Eq.~\eqref{eq:Qdot_T} and \eqref{eq:Qdot_T_improved} become quasi-independent of redshift. This means that most of the heating occurs in the $y$-era, while energy release during the $\mu$-$y$ transition era and the $\mu$-era is very small. From the observational point of view, this case is not as interesting, since it will be very hard to extract the primordial contribution to the $y$-parameter, given that a much larger distortion is created by reionization and structure formation and even from the damping of adiabatic modes (see Sect.~\ref{sec:late_release}). 
For $\nT=0.5$, the integral scales as $\simeq (1+z)$, so that $\id (Q/\rho_\gamma)/\id \ln z$ becomes roughly constant. Comparing with the level of heating for adiabatic modes, it is clear that the distortion should be about $\simeq 10^4$ times smaller. 
For $\nT=1$, we gain another factor of $\simeq (1+z)$ so that most energy causes a $\mu$-distortion. In this case, the heating rate in the $\mu$-era becomes comparable to the heating for standard adiabatic modes.

\vspace{-3mm}
\subsubsection{Expressions for simple estimates}
\label{sec:simp_express_Qdot}
The level of the heating due to tensors remains much smaller than for scalars unless a very blue tensor power spectrum is  assumed. To estimate the distortion analytically,  at $z\gtrsim 10^4$ we find 
\beal
\label{eq:Qdot_T_improved_approx}
\left.\frac{\id (Q/\rho_\gamma)}{\id \ln z}\right|_{\rm T}
&\approx
\begin{cases}
0.27 A_T \ln(k_{\rm cut}\eta)/(1 + z) & \text{for}\;\nT=0 
\\[1mm]
\frac{0.27 A_T}{\nT(1 + z)} \left(\frac{k_{\rm cut}}{k_0}\right)^{\nT}& \text{for}\, \nT\gtrsim 0.1 
\end{cases}
\end{align}
to represent the numerical results well. Here, the cutoff scale is determined by $k_{\rm cut}\eta \approx 0.2 (1 + z)$ and $k_{\rm cut}\approx \pot{4.5}{-7}(1+z)^2 \,\Mpc^{-1}$. These expressions were obtained with $\mathcal{T}_h(x)\approx 2[\cos(x)-\sin(x)/x]^2$ ($\equiv$ free solution) in Eq.~\eqref{eq:Qdot_T}, rescaling by $1.29/2\simeq 0.645$ to capture the overall reduction of the tensor amplitude by neutrino damping and then keeping the leading order term in $x_{\rm cut}=k_{\rm cut}\eta\gg 1$ only.

\vspace{-3mm}
\subsection{Energy release in the $y$-distortion era}
\label{sec:late_release}
For modes entering the horizon during the $y$-era ($z\lesssim 10^4$), we have to include modifications related to the transition from radiation to matter domination around $z\simeq \pot{3}{3}$. 
Even if generally $y$-distortion constraints are harder to interpret because a very large signal is produced at late times by structure formation and reionization, it is still interesting to ask how large the tensor contribution to the photon heating is. 
For modes that enter the horizon in the  matter-dominated era ($k<k_{\rm eq}\simeq 10^{-2} \,\Mpc^{-1}$), the free streaming damping from neutrinos can be neglected (they become dynamically subdominant). In this case, the approximate solution of the tensor transfer function reads \citep{Watanabe2006} $h'\simeq 3 j_2(k\eta)/\eta$, with $\eta=2 c/ (H a)\propto a^{-1/2}$ for matter domination. The partial heating rate from these large-scale modes thus is
\beal
\label{eq:Qdot_T_large_scales}
\left.\frac{\id (Q/\rho_\gamma)}{\id t}\right|_{\rm T, late}
&\approx
\frac{4}{45 \taudot}\frac{H^2}{4} \int_0^{k_{\rm eq}} 
\frac{k^2 \!\id k}{2\pi^2} P_T(k)\,\mathcal{T}_h(k \eta) 
\nonumber\\[1mm]
\mathcal{T}_h(x)&\approx 18\, j_2^2(x),
\end{align}
where we scaled out the leading term $\propto c^2/(a \eta)^2\approx H^2/4(\propto a^{-3})$ of the transfer function of $h'$. For $\nT=0$, we can evaluate the $k$-space integral, $I_{\rm mat}=\int_0^{k_{\rm eq}} \frac{k^2 \!\id k}{2\pi^2} P_T(k)\,\mathcal{T}_h(k \eta)$, numerically. If we instead use the transfer function for the radiation-dominated era, $\mathcal{T}_h(x)\approx 2 (k\eta)^2 j_1^2(k\eta)$, and compare the results, we find that typically $I_{\rm mat}/I_{\rm rad}\simeq 0.36-0.9$. For the heating rates shown in Fig.~\ref{fig:heat_rate_examples}, we assumed that the transfer function of $h'$ is given by the one for radiation domination. Since in the radiation-dominated era we have $c^2/(a \eta)^2\approx H^2 (\propto a^{-4})$, in Fig.~\ref{fig:heat_rate_examples} we overestimated the contributions from modes with $k<k_{\rm eq}$ at least by a factor of $I_{\rm rad}/(I_{\rm mat} /4) \simeq 5$. Because our numerical computations already show that the heating in the $y$-era remains very small (see Fig.~\ref{fig:heat_rate_examples} around $z\simeq 10^3-10^4$; although not shown, at $z\lesssim10^3$ we find the heating rate to drop sharply), we conclude that the late time heating always remains small and thus can be neglected.

\vspace{-0mm}
\subsection{Alternative derivation for the tensor heating rate}
\label{sec:consistency}
To check the consistency of our derivations, we can obtain the expression for the effective heating rate caused by tensors in another way, starting from the gravitational wave energy density, $\rho_{\rm gw}(z)$. 
The gravitational wave contribution to the energy density of the Universe can be written as\footnote{We obtained this expression from Eq.~(23) of \citet{Boyle2008a}, identifying the initial tensor power spectrum as $\Delta^2_h(k)=k^3 P_T(k)/(2\pi^2)$ and using $k^2 |h|^2 = {|h'|}^2$ with the transfer function $\mathcal{T}_h$ to relate the initial power to later time. We also included the tiny correction to the energy density caused by dissipation of energy in the photon fluid, Appendix~\ref{app:gamma_damp}, which energetically is not important for the tensor perturbations but it is the origin of the heating for photons.} \citep[e.g.,][]{Boyle2008a, Watanabe2006}
\beal
\label{eq:GW_rho}
\rho_{\rm gw}(z)
&\approx \rho_{\rm tot} \int_0^{k_{\rm cut}}\frac{k^2 \!\id k}{2\pi^2} \frac{P_T(k)}{12} \,\frac{\mathcal{T}_h(k\eta)}{2} \,\expf{-\Gamma_\gamma \eta},
\end{align}
where $k_{\rm cut}$ is a small-scale cutoff that will be discussed below. The tensor energy transfer function, $\mathcal{T}_h(k\eta)$, is given by Eq.~\eqref{eq:Qdot_T} and $\rho_{\rm tot}\approx \rho_\gamma/(1-R_\nu)$ denotes the total energy density of the Universe.

It is clear that without any energy exchange between gravity waves, neutrinos and photons, one has $\rho_{\rm gw}\propto a^{-4}$ in the radiation-dominated era. The time derivative $a^{-4} \!\id (a^4 \rho_{\rm gw})/\!\id t$ thus describes the real exchange of energy between different fluid components:
\beal
\label{eq:GW_rho_deriv}
\frac{\id (a^4 \rho_{\rm gw})}{a^4 \id t}
&\approx\rho_{\rm tot} \int_0^{k_{\rm cut}}\frac{k^2 \!\id k}{2\pi^2} \frac{P_T(k)}{12} \,\frac{\id}{\id t} \left(\frac{\mathcal{T}_h(k\eta)}{2} \,\expf{-\Gamma_\gamma \eta}\right).
\end{align}
The remaining time derivative describes the heating of the neutrino fluid, $\propto \mathcal{\dot T}_h$, and the heating of the photon fluid, proportional to
\beal\nonumber
\frac{\id}{\id t} \expf{-\Gamma_\gamma \eta}=-\frac{32 H^2 (1-R_\nu)}{15\taudot} \, \expf{-\Gamma_\gamma \eta}, 
\end{align}
where we used the definition of $\Gamma_\gamma$ given in Appendix~\ref{app:gamma_damp}. Thus, the transfer of energy from tensors to the photon field is given by
\beal
\label{eq:GW_rho_deriv_gamma}
\left.\frac{\id (a^4 \rho_{\rm gw})}{a^4 \id t}\right|_{\gamma}
&\approx\rho_{\rm tot} \int_0^{k_{\rm cut}}\frac{k^2 \!\id k}{2\pi^2} \frac{P_T(k)}{12} \,\frac{\mathcal{T}_h(k\eta)}{2} 
\, \frac{\id}{\id t} \expf{-\Gamma_\gamma \eta}
\nonumber\\
&=-\frac{32 H^2\rho_{\rm tot} (1-R_\nu)}{15\taudot} \int_0^{k_{\rm cut}}\frac{k^2 \!\id k}{2\pi^2} \frac{P_T(k)}{12} 
\,\frac{\mathcal{T}_h(k\eta)}{2} \,\expf{-\Gamma_\gamma \eta}
\nonumber\\[1mm]
&=-\frac{4 H^2}{45\taudot} \, \rho_{\gamma}
\int_0^{k_{\rm cut}}\frac{k^2 \!\id k}{2\pi^2} \,P_T(k) 
\,\mathcal{T}_h(k\eta) \,\expf{-\Gamma_\gamma \eta}.
\end{align}
Comparing this with Eq.~\eqref{eq:Qdot_T}, we can confirm our expression for the effective heating rate of photons by tensors. For the shear viscosity from photons, transfer effects were neglected for Eq.~\eqref{eq:Qdot_T}. These lead to a scale-dependent correction of the damping factor, $\Gamma^\ast_\gamma(k, \eta)$, that at different level of precision can be deduced from Eq.~\eqref{eq:Qdot_T_improved}, \eqref{eq:T_Qdot_l2} or \eqref{eq:quasi-exact}. Also, in principle additional changes due to modifications of the effective number of relativistic degrees of freedom can be accounted for, which leads to modulation of the tensor power relative to the $\rho_{\rm gw}\propto a^{-4}$ scaling, but the basic conclusion does not change.

\vspace{-5mm}
\section{Results for $\mu$-distortion from tensors}
\label{sec:mu_res}
Given the heating rate from tensor perturbations, we can estimate the amplitude of the $\mu$-distortion using \citep[e.g.,][]{Hu1993}
\beal
\label{eq:mu_approx}
\mu&\approx 1.4\int_{z_{\mu,y}}^{\infty}
\left.\frac{\id (Q/\rho_\gamma)}{\id z}\right|_{\rm T} \expf{-(z/\zmudc)^{5/2}}\id z,
\end{align}
with $z_{\mu,y}\simeq \pot{5}{4}$ and $\zmudc\simeq \pot{2}{6}$. Here, $\mathcal{J}(z)=\expf{-(z/\zmudc)^{5/2}}$ gives a simple approximation of the distortion visibility function, which accounts for the efficiency of thermalization at early times. Corrections to the shape of the spectral distortion caused by dissipation of tensor perturbations in the $\mu-y$ transition era ($10^4 \lesssim z\lesssim \pot{3}{5}$) can be included using the Green's function method of the {\sc CosmoTherm}\footnote{Available at \url{www.Chluba.de/CosmoTherm}} software package \citep{Chluba2011therm, Chluba2013Green}, but for the purpose of this work, Eq.~\eqref{eq:mu_approx} is sufficient.

\begin{figure}
\centering
\includegraphics[width=1.02\columnwidth]{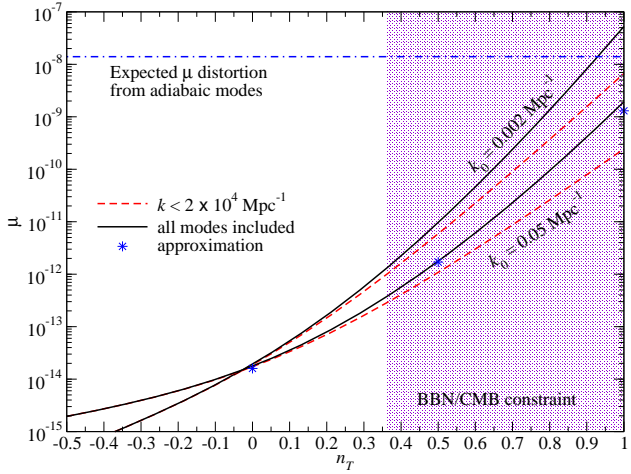}
\caption{Resulting $\mu$-parameter from heating due to tensor perturbations. The two groups are for $\{A_T, k_0 \}=\{\pot{2.4}{-10}, 0.002\,\Mpc^{-1}\}$ and $\{\pot{2.2}{-10}, 0.05\,\Mpc^{-1}\}$. We used Eq.~\eqref{eq:quasi-exact} to compute the heating rate, but for the red dashed line we only included modes with $k\leq \pot{2}{4}\,\Mpc^{-1}$. The stars show the result obtained with  approximation Eq.~\eqref{eq:Qdot_T_improved_approx}. For the simplest parametrizations of the primordial tensor power spectrum, the shaded region is in tension with BBN/CMB constraints \citep[see][and Sect.~\ref{sec:mu_res} for more details]{Smith2006, Boyle2008b}}.
\label{fig:mu_plot}
\end{figure}

For $k_0=0.05\,\Mpc$, with the approximation Eq.~\eqref{eq:Qdot_T_improved_approx} for the tensor heating rate, we find $\mu\approx \{\pot{7.3}{-5}, \pot{7.8}{-3}, 5.8\}\,A_T$ for $\nT=\{0, 0.5, 1\}$, respectively. Thus, with $A_T\simeq 0.1 A_\zeta\simeq \pot{2.2}{-10}$, we have a distortion $\mu\approx \{\pot{1.6}{-14}, \pot{1.7}{-12}, \pot{1.3}{-9}\}$. For $\nT\lesssim 1$, this agrees to within $\simeq 10\%-30\%$ with our more detailed calculation (see Fig.~\ref{fig:mu_plot}). Generally, our numerical results show that for nearly scale invariant tensor power spectra, the $\mu$-distortion remains six orders of magnitudes smaller than for the dissipation of adiabatic modes, which for standard curvature power spectrum with $A_\zeta=\pot{2.2}{-9}$ at pivot scale $k_0=0.05\,\Mpc$ and $\nS=0.96$ gives $\mu_\zeta\simeq \pot{1.4}{-8}$ \citep{Chluba2012}. The adiabatic signal is just at the detection limit of PIXIE \citep{Kogut2011PIXIE}, showing that a detection of the tensor contribution is very futuristic. For blue power spectra, the distortion can become comparable to the signal caused by adiabatic modes. However, in this case, constraints on tensors from CMB and big bang nucleosynthesis (BBN) become important \citep{Smith2006}, limiting $\nT<0.36$ for $r\simeq 0.1$ and the simplest parametrization for the tensor power spectrum \citep{Boyle2008b}. 
These constraints are derived using several approximations, which can affect the constraint on $\nT$ significantly. Constraints on extended models have recently been discussed by \citet{Kuroyanagi2014}.
Overall, the distortion signal from tensors is still expected to be much smaller than for adiabatic modes (see Fig.~\ref{fig:mu_plot}).

\vspace{0mm}
\subsection{Comparing with Ota et al.}
\label{sec:comp_Ota}
Our conclusions from the previous section are in broad agreement with those of \citet{Ota2014}. To compare more directly, we change the power spectrum parameters to $k_0=0.002\,\Mpc$ and $A_T=\pot{2.4}{-10}$ and introduce a hard small-scale cutoff at $k_{\rm cut}=\pot{2}{4}\,\Mpc^{-1}$. Like \citet{Ota2014}, here, we also neglect the heating from multipoles with $\ell>2$ and extra polarization terms, although these add a significant correction to the heating rate in the free streaming regime (see Fig.~\ref{fig:heating_one_mode_comp}).
Numerically integrating Eq.~\eqref{eq:Qdot_T_improved} with Eq.~\eqref{eq:mu_approx}, we find $\mu\approx \{\pot{1.8}{-14}, \pot{6.0}{-9}\}$ for $\nT=\{0, 1\}$. This is about $10\%-20\%$ smaller than the values reported in their paper, $\mu_{\rm Ota}\approx \{\pot{2.2}{-14}, \pot{7}{-9}\}$ for\footnote{We refer to the values quoted in arXiv:1406.0451 v2, which are slightly lower than for v1.} $r=0.1$. A part of this difference can be explained by adding the other terms for $\ell=2$, Eq.~\eqref{eq:T_Qdot_l2}, which are neglected for the approximation Eq.~\eqref{eq:Qdot_T_improved} and then give $\mu\approx \{\pot{1.9}{-14}, \pot{6.3}{-9}\}$. We find, however, that using Eq.~\eqref{eq:T_Qdot_l2} overestimates the $\ell=2$ contribution by about $\simeq 5\%-10\%$, since the corrections to the $\ell=2$ transfer functions caused by higher $\ell$ multipoles are not included, but lead to an additional suppression of the $\ell=2$ terms. Thus, in particular for $\nT=0$, the difference remains comparable to $\simeq 20\%$.

\begin{figure}
\centering
\includegraphics[width=1.02\columnwidth]{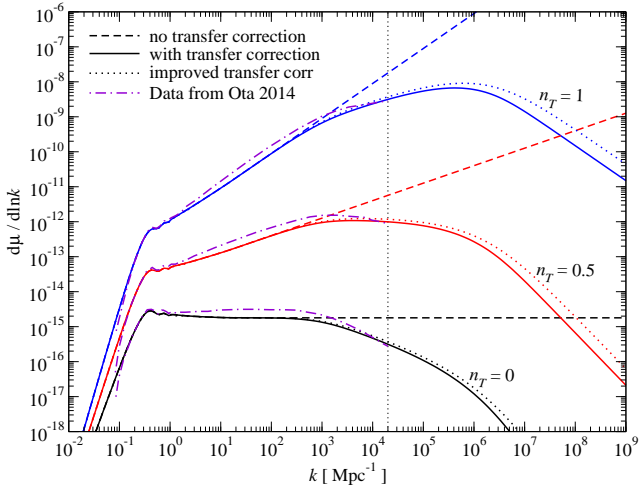}
\caption{Differential contribution to the $\mu$-distortion from different scales. Transfer effects introduce a cutoff at very small scales. The dotted vertical line indicates the position of the cutoff used by \citet{Ota2014}, while the dash-dotted lines are the data taken from their Fig. 2 (and divided by $2$ to convert to $r=0.1$). See Sect.~\ref{sec:comp_Ota} text for more detailed explanation.}
\label{fig:dmu_dlnk}
\end{figure}

To understand the remaining difference a little better, in Fig.~\ref{fig:dmu_dlnk} we show the digitized points (purple, dash-dotted) for $\id \mu/\id \ln k$ taken from Fig.~2 of \citet{Ota2014} in comparison with our numerical results. For the solid lines, we used Eq.~\eqref{eq:Qdot_T_improved} for the heating rate, while the dotted lines were computed with Eq.~\eqref{eq:quasi-exact} for the photon transfer function. For illustration, we also show the result for $\id \mu/\id \ln k$, when neglecting any photon transfer effects (dashed lines), which emphasizes the importance of free streaming effects.
At the largest scales ($k\simeq 0.3\,\Mpc^{-1}$), our curves for $\id \mu/\id \ln k$ practically coincide, although we find slightly larger contributions at $k\lesssim 0.1\,\Mpc^{-1}$. However, at smaller scales, the curves of \citet{Ota2014} are roughly $1.5$ times larger than ours. 
\citet{Ota2014} used the numerical output from the CLASS code \citep{CLASSCODE, CLASSII, Tram2013} to obtain the transfer functions. The effect of neutrino damping was only included to CLASS recently (i.e., version 2.2; private communication, Lesgourgues). We find that after neglecting the damping effect of neutrinos our curves practically agree. Nevertheless, these corrections do not change any of the main conclusions.

We do, however, find that modes at $k\gtrsim\pot{2}{4}\,\Mpc^{-1}$, which were neglected by \citet{Ota2014}, contribute significantly to the heating, in particular for blue tensor power spectra. Including all modes relevant at smaller scales, for $k_0=0.002\,\Mpc$ and $A_T=\pot{2.4}{-10}$, we find $\mu\approx \{\pot{1.9}{-14}, \pot{5.3}{-8}\}$. Due to the logarithmic dependence of the heating rate on the small-scale cutoff [cf., Eq.~\eqref{eq:Qdot_T_improved_approx}], for $\nT=0$ this did not make much of a difference. However, for $\nT\simeq 1$, the distortion is underestimated roughly seven times when neglecting modes at $k>\pot{2}{4}\,\Mpc^{-1}$ (see Fig.~\ref{fig:mu_plot}). This becomes apparent when looking at the differential contribution to $\mu$ as a function of scale (Fig.~\ref{fig:dmu_dlnk}). For $\nT=1$, even scales up to $k\simeq 10^8\,\Mpc^{-1}$ contribute significantly to the value of $\mu$, which again emphasizes that for tensors spectral distortions are sensitive to much smaller scales than for scalars. We mention, however, that even our results need refinements in this regime, since we neglected several effects that modify the tensor power spectrum at small scales by $\simeq 10\%-30\%$ [see discussion in Sect.~\ref{sec:Tensor_TC_I}].

\vspace{0mm}
\subsection{Window function in $k$-space for scalar and tensor modes}
\label{sec:window}
\begin{figure}
\centering
\includegraphics[width=1.03\columnwidth]{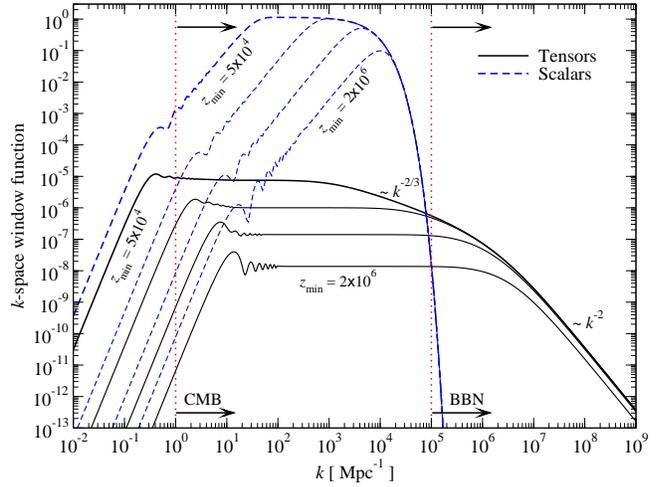}
\caption{Comparison of the $k$-space window functions for adiabatic modes and tensors. To illustrate the redshift dependence of the heating rate, we also vary the lower redshift in the integrals, Eq.~\eqref{eq:Window_i}. In each group, we used $z_{\rm min}=\{\pot{5}{4}(\equiv z_{\mu,y}), \pot{3}{5}, 10^6, \pot{2}{6}\}$, respectively. We also indicated the scales that are relevant for the integrated constraints of the tensor power from the CMB damping tail and BBN measurements \citep[see][]{Smith2006}.}
\label{fig:window}
\end{figure}

Another way to illustrate the dependence of the distortion signal on scale is to introduce $k$-space window functions that determine the contributions to the $\mu$-distortion from different modes. A similar procedure was used by \citet{Chluba2012inflaton} and \cite{Chluba2013iso} to compute the signals for adiabatic and isocurvature modes. 
The window function can be directly obtained from the definition of the effective heating rates, Eq.~\eqref{Qdot_Sc} and \eqref{eq:quasi-exact}, and the approximation for $\mu$, Eq.~\eqref{eq:mu_approx}. With this, for scalars and tensors we may write 
\beal
\label{eq:mu_Window}
\mu_i\approx \int_0^\infty \frac{k^2 \!\id k}{2\pi^2} P_i(k) W_i(k),
\end{align}
where $i = \{\zeta, T\}$. The window functions are 
\bsub
\label{eq:Window_i}
\beal
\label{eq:Window_i_a}
W_\zeta(k)&\approx  1.4\int_{z_{\mu,y}}^{\infty} 
 \frac{32 c^2 k^2}{45 a^2 \taudot} D^2 \sin^2(k \rs) \,\expf{-2k^2/\kD^2}\,\expf{-(z/\zmudc)^{5/2}}\id z
 \\
\label{eq:Window_i_b}
W_T(k)&\approx  1.4\int_{z_{\mu,y}}^{\infty} 
 \frac{4H^2}{45 \taudot} \,\mathcal{T}_h(k\eta) \,\mathcal{T}_\Theta(k/\tau') \,\expf{-\Gamma^\ast_\gamma \eta} \,\expf{-(z/\zmudc)^{5/2}}\id z.
\end{align}
\esub
The results for $W_i$ are shown in Fig.~\ref{fig:window}. For adiabatic perturbations, most of the contributions to the value of $\mu$ arise from scales ${\rm few}\,\Mpc^{-1}\lesssim k \lesssim\pot{\rm few}{4}\,\Mpc^{-1}$, while for tensor perturbations modes with wavenumbers $0.1\,\Mpc^{-1}\lesssim k \lesssim\pot{\rm few}{5}\,\Mpc^{-1}$ contribute significantly for nearly scale invariant power spectra. As explained above, this is due to the fact that for adiabatic modes the damping by photon diffusion plays an important role, while for tensors free streaming is relevant. 
We can furthermore see that for adiabatic perturbations, the heating at early times is dominated by the smallest scales, while for tensors the heating in different epochs is less scale dependent.

From Fig.~\ref{fig:window}, we can also conclude that CMB spectral distortion measurements from COBE/FIRAS for individual modes do not give any stringent constraint on the tensor power spectrum at small scales. Directly translating $\mu\lesssim \pot{9}{-5}$ ($95\%$ c.l.) yields $k^3 P_T(k)/(2\pi^2)\lesssim 10$ at $0.45\,\Mpc^{-1}\lesssim k \lesssim 250\,\Mpc^{-1}$ and even weaker otherwise. Clearly, in this case a simple linear analysis is questionable.
For adiabatic modes, we have the much stronger limit $k^3 P_\zeta(k)/(2\pi^2)\lesssim \pot{8}{-5}$ at $50\,\Mpc^{-1}\lesssim k \lesssim 10^3\,\Mpc^{-1}$ \citep[see,][]{Chluba2012inflaton}. Even for the integrated power the limit remains extremely weak, only giving $\int k^2\id k\, P_T(k)/(2\pi^2)\lesssim 1$ at $0.45\,\Mpc^{-1}\lesssim k \lesssim 250\,\Mpc^{-1}$ from COBE/FIRAS. With a PIXIE-like experiment, this could tighten by a factor of $\simeq 10^3-10^4$, providing a constraint on a part of the tensor power spectrum that is complementary (although weaker) to, e.g., pulsar timing measurements, CMB and future gravitational wave observatories \citep[e.g.,][]{Smith2006, Boyle2008b}. In principle, this could help to rule out very non-standard early-universe scenarios.

\vspace{-0mm}
\section{Conclusions}
\label{sec:conclusions}
We obtained general expressions for the effective heating rate caused by scalar, vector and tensor perturbations (Sect.~\ref{sec:heating_general}). These expressions include previously neglected terms from polarization states and contributions from higher multipoles, which become noticeable when the tight coupling approximation breaks down. We explicitly confirmed that only scattering terms are relevant for the dissipation process of scalar, vector and tensor perturbations (Appendix~\ref{app:proof}).
We furthermore showed that the heating rate due to tensors can be approximated very well using tight coupling solutions with additional radiative transfer corrections in the quasi-free streaming regime [see Eq.~\eqref{eq:Qdot_T_improved}]. The required photon transfer functions can be derived analytically, as we explain in Appendix~\ref{app:ana_sol_env}. These expressions represent both the amplitude and phase of the photon transfer functions for $\ell=2$ very well.
Using energetics arguments, we also directly linked the photon heating term to the loss of energy from the tensor perturbations (see Sect.~\ref{sec:consistency}), confirming the normalization of our analytic expressions for the heating rate.

Without additional radiative transfer corrections, the heating rate from tensors is practically scale independent. However, scale dependence is introduced due to free streaming. This is in stark contrast to adiabatic perturbations, for which the relevant scales is related to photon diffusion. Since the free streaming scales is smaller than the damping scale for adiabatic modes, spectral distortions probe tensor perturbations to significantly smaller scales. In particular, we find that for scale invariant tensor power spectrum, distortions in the $\mu$-era are sourced by tensor perturbations modes with wavenumbers $0.1\,\Mpc^{-1}\lesssim k \lesssim\pot{\rm few}{5}\,\Mpc^{-1}$ (see Fig.~\ref{fig:window}). Even smaller scales become important for blue tensor power spectra, since the $k$-space distortion window function only decays as a power law $\simeq k^{-2}$ (instead of exponentially as for adiabatic perturbations). The small-scale contributions were previously ignored, but can affect the distortion amplitude significantly (see Fig.~\ref{fig:mu_plot}). We also argue that the heating from tensors caused during the $y$-era remains subdominant (see Sect.~\ref{sec:late_release}).

For scale independent tensor power spectra with tensor amplitude $A_T=\pot{2.2}{-10}$ at pivot scale $k_0=0.05\,\Mpc$ we find a distortion $\mu\simeq \pot{1.8}{-14}$, while for $A_T=\pot{2.4}{-10}$ at pivot scale $k_0=0.002\,\Mpc$, we have $\mu\simeq \pot{1.9}{-14}$. This is some six orders of magnitudes smaller than for adiabatic modes and thus extremely challenging to detect. 
For very blue tensor power spectra with $\nT\simeq 1$, we obtain $\mu\simeq \pot{1.9}{-9}$, while using $A_T=\pot{2.4}{-10}$ at pivot scale $k_0=0.002\,\Mpc$, we find $\mu\simeq \pot{5.3}{-8}$.
This signal is comparable to the one for adiabatic modes in the standard inflation scenario, $\mu\simeq \pot{1.4}{-8}$ \citep{Chluba2012}; however, constraints from BBN limit $\nT<0.36$ for the simplest models \citep{Boyle2008b}, so that $\nT\simeq 1$ is already in tension with this. It is, however, important to emphasize that these constraints make certain assumptions about the standard number of relativistic degrees of freedom and the scale dependence of the tensor power spectrum \citep[see][for discussion of more general cases]{Kuroyanagi2014}, so that independent constraints from CMB spectral distortions are still valuable. 
For $\nT\simeq 0.36$, we find $\mu\simeq \pot{3.7}{-13}$ [$(A_T, k_0)=(\pot{2.2}{-10},0.05\,\Mpc)$] and $\mu\simeq \pot{1.3}{-12}$ [$(A_T, k_0)=(\pot{2.4}{-10},0.002\,\Mpc)]$, showing that overall the distortion caused by tensor perturbations is expected to be more than four orders of magnitudes smaller than from adiabatic modes. 
Similarly, the amount of entropy production due to tensor damping in the temperature era ($z\gtrsim \pot{2}{6}$) will be many orders of magnitudes smaller than for scalars \citep[for discussion of this process see,][]{Jeong2014}.
Our results are generally in good agreement with those of \citet{Ota2014}, although here we included several additional effects, such as the damping of neutrinos, extra heating from small scales ($k\gtrsim \pot{2}{4}\,\Mpc^{-1}$), contributions from $\ell>2$ and additional polarization corrections. For more details, see Sect.~\ref{sec:comp_Ota}.

For the future, it is possible to improve our estimates using more precise computations for the shape of the tensor power spectrum \citep[e.g.,][]{Boyle2008b, Watanabe2006}. Overall, we expect these corrections to affect the results at the level of $\simeq 20\%-30\%$. It could also be interesting to study the prospects of constraining variations of the $\mu$ and $y$ distortion signal introduced by spatial variations of the tensor power across the sky. However, from our computations, it is clear that a signal could only be detectable if the small-scale tensor power is very large and modulated by large-scale modes due to non-Gaussianity in the squeezed limit. This requires very non-standard early-universe models and thus is left for future explorations.

\small

\vspace{0mm}
\section*{Acknowledgments}
The authors thank Julien Lesgourgues, Wayne Hu and Atsuhisa Ota for helpful discussions and comments. 
JC, LD and MK are supported by NSF Grant No. 0244990 and the John Templeton Foundation. DG is funded at the University of Chicago by a National Science Foundation Astronomy and Astrophysics Postdoctoral Fellowship under Award NO. AST-1302856. MA is supported by a Kavli Fellowship at the University of Cambridge.

\begin{appendix}

\vspace{-0mm}
\section{Generalizing the superposition of blackbodies with linear polarization}
\label{app:sup_gen_deriv}
We can generalize Eq.~\eqref{eq:sup_bb} to partially polarized light (linear polarization only) using a density matrix representation for the individual polarization states. Writing the two polarizations independently, we may introduce the $2\times 2$ occupation matrix
\beal
\mathcal{N}_{\rm unpol}=\matpol{\nbb(x)}{0}{0}{\nbb(x)}
\end{align}
and the total occupation number $n=\frac{1}{2} {\rm Tr}(\mathcal{N})=\nbb(x)$ (averaged over both polarization states). For partially polarized light, the occupation number in the two polarization directions differs. Assuming that the polarization is aligned with one of the directions of the polarization basis we thus have
\beal
\mathcal{N}=\matpol{\nbb(x/[1+\Theta_\parallel])}{0}{0}{\nbb(x/[1+\Theta_\perp])},
\end{align}
where the two blackbodies have temperatures $T_i=T(1+\Theta_i)$. We can now rewrite the occupation matrix as 
\beal
\mathcal{N}=\frac{n_\parallel+n_\perp}{2}\!\matpol{1}{0}{0}{1}+\frac{n_\parallel-n_\perp}{2}\!\matpol{1}{0}{0}{-1}
=n_I {\mathbf{1}}+n_Q{\boldsymbol\sigma}_3.
\end{align}
Here, ${\boldsymbol\sigma}_i$ denote the Pauli-spin matrices. Then, the two occupations numbers $n_I=(n_\parallel+n_\perp)/2$ and $n_Q=(n_\parallel-n_\perp)/2$ can be expressed as
\beal
n_I&\approx \nbb(\bar x)+G(\bar x) \frac{\bar \Theta_\parallel +\bar\Theta_\perp+\bar \Theta^2_\parallel +\bar\Theta^2_\perp}{2}
+Y_{\rm SZ}(\bar x) \frac{\bar \Theta^2_\parallel +\bar\Theta^2_\perp}{4}
\nonumber\\[-0.5mm]
n_Q&\approx  \qquad\quad\,\, G(\bar x) \frac{\bar \Theta_\parallel -\bar\Theta_\perp+\bar \Theta^2_\parallel - \bar\Theta^2_\perp}{2}
+Y_{\rm SZ}(\bar x) \frac{\bar \Theta^2_\parallel - \bar\Theta^2_\perp}{4}
\end{align}
when expanding around a blackbody at temperature $ \bar T$ (e.g., defined by an all-sky average). Here, $\bar x = x\, T/\bar T=h\nu/k \bar T$ and we have $T_\parallel=\bar T (1+\bar\Theta_\parallel)$ and $T_\perp=\bar T (1+\bar\Theta_\perp)$. We identify the usual intensity and Stokes $Q$ temperature perturbations, $\Theta_I=(\bar \Theta_\parallel +\bar\Theta_\perp)/2$ and $\Theta_Q=(\bar \Theta_\parallel -\bar\Theta_\perp)/2$, so that
\beal
n_I&\approx \nbb(\bar x)+G(\bar x) 
\left( \Theta_I + \Theta^2_I + \Theta^2_Q\right) +Y_{\rm SZ}(\bar x) \frac{\Theta^2_I + \Theta^2_Q}{2}
\nonumber\\
n_Q&\approx  \qquad\quad\,\, G(\bar x) \left(\Theta_Q+ 2 \Theta_I \Theta_Q \right)
+Y_{\rm SZ}(\bar x) \Theta_I \Theta_Q.
\end{align}
This shows that the energy distribution of the Stokes $Q$ parameter reflects that of a temperature perturbation [$\propto G(x)$] with a $y$-distortion at second order. We furthermore see that at second order in $\Theta_i$, $Q$ contributes to the energy distribution of $I$.

For general polarization state, we also need to consider non-zero occupation of the Stokes $U$ parameter, i.e, $n_U=\frac{1}{2} {\rm Tr}(\mathcal{N} {\boldsymbol\sigma}_1)$. This can be obtained by rotating the polarization basis. The generalization thus is
\beal
\mathcal{N}&=n_I\,{\mathbf{1}}+n_Q\,{\boldsymbol\sigma}_3+n_U\,{\boldsymbol\sigma}_1
\nonumber\\[-2.8mm]
n_I&\approx \nbb(\bar x)+G(\bar x) 
\left( \Theta_I + \Theta^2_I + \Theta^2_Q+\Theta^2_U\right) +Y_{\rm SZ}(\bar x) \frac{\Theta^2_I + \Theta^2_Q+\Theta^2_U}{2}
\nonumber\\[-1.0mm]
n_Q&\approx  \qquad\quad\,\, G(\bar x) \left(\Theta_Q+ 2 \Theta_I \Theta_Q \right)
+Y_{\rm SZ}(\bar x) \Theta_I \Theta_Q
\nonumber\\
\label{eq:occ_IQU}
n_U&\approx  \qquad\quad\,\, G(\bar x) \left(\Theta_U+ 2 \Theta_I \Theta_U \right)
+Y_{\rm SZ}(\bar x) \Theta_I \Theta_U.
\end{align}
The energy distribution of both $Q$ and $U$ have the same form and both contribute to the spectrum of Stokes $I$.

\vspace{-5mm}
\section{Effective heating term with vector and tensor perturbations}
\label{app:proof}
To understand all contributions to the spectral distortion evolution caused by the Liouville operator, we start from the photon Boltzmann equation 
\beal
\label{eq:BE_gen}
\frac{\id f}{\id \eta}= \frac{\partial f}{\partial \eta} + \frac{\partial f}{\partial x^i} \frac{\id x^i}{\id\eta} + \frac{\partial f}{\partial p}\frac{\id p}{\id\eta} + \frac{\partial f}{\partial n^i} \frac{\id n^i}{\id\eta}
\end{align}
with definitions from \citet{Bartolo2006}. The photon phase space distribution at different perturbation order is
\bsub
\beal
f^{(0)}&=f_{\rm bb}+\Delta f^{(0)}
\\
f^{(1)}&=\Delta f^{(1)} + f_{T} \Theta^{(1)}
\\[-0.8mm]
f^{(2)}&=\Delta f^{(2)} + f_{T} \left([\Theta^{(1)}]^2+\Theta^{(2)}\right)+\frac{1}{2}\, f_{y}\, [\Theta^{(1)}]^2,
\end{align}
\esub
where $\Delta f^{(k)}$ describe spectral distortions, $f_T$ temperature terms and $f_{y}$ the $y$-distortion contribution from the superposition of blackbodies. Since in the tight coupling limit polarization states do not contribute to the heating, we focus on the distribution function summed over polarization only ($[\Theta^{(1)}]^2=[\Theta^{(1)}]^2_I+[\Theta^{(1)}]^2_Q+[\Theta^{(1)}]^2_U$, but we use short notation here).
If we consider only terms that have a $y$-type dependence, then in Eq.~\eqref{eq:BE_gen} we need to keep the $y$-part of all terms $\propto f^{(2)}$ using $\id x^{i, (0)}/\id\eta=n^i$, $\id p^{(0)}/\id\eta=-\mathcal{H} p$ and $\id n^{i,(0)}/\id\eta=0$. The derivative $\partial f/\partial p$ furthermore creates a $y$-term from occurrences of $f^{(1)}$, or explicitly $p \,\partial f^{(1)}/\partial p \rightarrow - f_{y} \Theta^{(1)}$ and we need $\id \ln p^{(1)}/\id\eta=-{\Phi'}^{(1)}+\partial_{x^i}(\Phi^{(1)}+\Psi^{(1)}-\omega^{(1)}_i n^i -\frac{1}{2} h^{(1)}_{ij}n^i n^j)$, where $\Phi, \Psi, \omega_i$ and $h_{ij}$ describe the metric perturbations in the Poisson gauge. 

At first order, no distortion is created unless anisotropic energy release occurs. This is because perturbations only create fluctuation with thermal spectrum and the scattering physics always pushes electrons locally into equilibrium with the monopole of the photon field, so that no distortion is sourced \citep{Chluba2012}, i.e. $\Delta f^{(1)}=0$. Then, all terms from Eq.~\eqref{eq:BE_gen} that relate to distortions are 
\beal
\label{eq:BE_gen_dist}
\frac{\id f}{\id \eta} \rightarrow 
D_\eta \left(\Delta f^{(2)} + \frac{1}{2} f_y [\Theta^{(1)}]^2\right) - f_{y} \Theta^{(1)} \frac{1}{p}\!\frac{\id p^{(1)}}{\id\eta},
\end{align}
where $D_\eta=\partial_\eta + n^i \partial_{x^i}$. The Hubble term was absorbed by transforming from $p\rightarrow x=h\nu/kT_\gamma$ with $T_\gamma\propto (1+z)$. Similarly, one can obtain an equation that includes $\Theta^{(2)}$ to describe the change of the average CMB monopole, but we omit these terms here.

For the temperature fluctuations at first order in perturbation theory, we only have
\beal
\label{eq:TP_gen_dist}
D_\eta \Theta^{(1)} - \frac{1}{p}\!\frac{\id p^{(1)}}{\id\eta} = {\rm C}^{(1)}[f]
\end{align}
again after absorbing the redshifting term $\propto \mathcal{H}$. Here, we introduced the collision term, ${\rm C}^{(1)}[f]$, due to scattering of first-order temperature perturbations. Using $D_\eta[\Theta^{(1)}]^2=2 \Theta^{(1)} D_\eta \Theta^{(1)}$, with Eq.~\eqref{eq:BE_gen_dist} and \eqref{eq:TP_gen_dist}, we find
\beal
\label{eq:BE_gen_dist_cancelation}
D_\eta \Delta f^{(2)} + f_y \Theta^{(1)} {\rm C}^{(1)}[f]={\rm C}^{(2)}[f],
\end{align}
where the second-order collision term was introduced. For pure temperature perturbations it, was shown that for the average spectrum of the CMB, ${\rm C}^{(2)}[f]$ sources one additional $y$-type term $\propto -\beta[\Theta^{(1)}_{1}-\beta/3]$ with $\beta=\varv/c$ \citep{Chluba2012}. There terms were caused by second-order scattering and give a fully gauge-independent expression for the distortion source terms. Similar terms are expected to appear when including the scattering process for polarized radiation. Again, this should fix the gauge dependence in the components of $\Theta^{(1)}_{1m}$ for $m\neq 0$. These terms should not affect the result in the tight coupling limit so we do not go into more detail. We thus have proven the statement that only terms related to the scattering need to be included for the computation of the distortion source, while metric terms do not directly source a distortion. These are, however, expected to affect the average monopole temperature term, an effect we neglect here.

\vspace{0mm}
\section{Simplifying the heating integral}
\label{app:integrals_long}
To compute $\big<\Theta_I \dot \Theta_I\big>$, we first evaluate the integral over photon directions, $\vek{n}$. Here, it is important to emphasize that for the time derivatives $\dot \Theta_I$ we only need to account for the scattering terms.
From Eq.~\eqref{eq:Fourier_conventions}, we have
\beal
\int \Theta_I\dot \Theta_I \frac{\id^2 n}{4\pi}&=\int \!\frac{\id^3 k}{(2\pi)^3}\frac{\id^3 k'}{(2\pi)^3}\,\expf{i \scriptsize \vek{x}\cdot(\vek{k}+\vek{k}')}
\!\sum_{\ell,\ell'} \!\sum_{m, m'} 
\frac{(-i)^{\ell+\ell'} }{\sqrt{(2\ell+1)(2\ell'+1)}} 
\nonumber\\
&\qquad\times\Theta_\ell^{(m)}(\vek{k})\,\dot\Theta_{\ell'}^{(m')}(\vek{k}') 
\int {}_0 Y_{\ell m}(\vek{n}) \,{}_0 Y_{\ell' m'}(\vek{n}) \id^2 n 
\nonumber\\
&=\int \!\frac{\id^3 k}{(2\pi)^3}\frac{\id^3 k'}{(2\pi)^3}\,\expf{i \scriptsize \vek{x}\cdot(\vek{k}+\vek{k}')}
\!\sum_{\ell,\ell'} \!\sum_{m, m'} 
\frac{(-i)^{\ell+\ell'} }{\sqrt{(2\ell+1)(2\ell'+1)}} 
\nonumber\\
&\qquad\times\Theta_\ell^{(m)}(\vek{k})\,\dot\Theta_{\ell'}^{(m')}(\vek{k}') (-1)^m \delta_{\ell,\ell'}\delta_{m,-m'}
\nonumber\\
&=\int \!\frac{\id^3 k}{(2\pi)^3}\frac{\id^3 k'}{(2\pi)^3}\,\expf{i \scriptsize \vek{x}\cdot(\vek{k}+\vek{k}')}
\sum_{\ell} \!\sum_{m=-2}^{m=2}\frac{ \Theta_\ell^{(m)}(\vek{k})\,(-1)^{\ell+m} \dot\Theta_{\ell}^{(-m)}(\vek{k}')}{(2\ell+1)} 
\nonumber\\ \nonumber
&=\int \!\frac{\id^3 k}{(2\pi)^3}\frac{\id^3 k'}{(2\pi)^3}\,\expf{i \scriptsize \vek{x}\cdot(\vek{k}-\vek{k}')}
\sum_{\ell} \!\sum_{m=-2}^{m=2}\frac{\Theta_\ell^{(m)}(\vek{k})\,\dot\Theta_{\ell}^{(m)}(\vek{k}')}{(2\ell+1)},
\end{align}
where in the last step we used $\Theta_{\ell}^{(m)}(\vek{k})=(-1)^{\ell+m} \Theta_{\ell}^{(-m)}(-\vek{k})$ to ensure that $\Theta_I$ is real and then redefined the integration variable $\vek{k}'\rightarrow -\vek{k}'$. 

We now use the transfer function to relate $\Theta_{\ell}^{(m)}$ at some time to the initial perturbations $\delta^{(m)}(\vek{k})$ using the replacement $\Theta_{\ell}^{(m)}(\vek{k})\rightarrow \Theta_{\ell}^{(m)}(\vek{k})\,\delta^{(m)}(\vek{k})$. The spatial average can then be carried out as ensemble average over universes, which ensures $\big<\delta^{(m)}(\vek{k})\,\delta^{(m)}(\vek{k}')\big>=(2\pi)^3 \delta(\vek{k}-\vek{k}') P^{(m)}_i(k)$ when assuming statistical isotropy. Here, $P^{(m)}_i(k)$ is the initial power spectrum of the perturbation variable with respect to which the transfer functions are defined. For scalar perturbations ($m=0$), assuming adiabatic perturbations, the curvature power spectrum, $P_\zeta(k)$, is used to set up the initial conditions, while for tensors ($m=\pm 2$) the transfer functions are defined with respect to the amplitude of $h$. The connection of $P^{(m)}_i(k)$ and $P_h(k)$ with the usual tensor power spectrum $P_T(k)$ will be clarified in Sect.~\eqref{sec:Tensor_TC_I}.  We thus find
\beal
\big< \Theta_I\dot \Theta_I \big>
&=\int \!\frac{\id^3 k}{(2\pi)^3}\frac{\id^3 k'}{(2\pi)^3}\,\expf{i \scriptsize \vek{x}\cdot(\vek{k}-\vek{k}')}
\sum_{\ell} \!\sum_{m=-2}^{m=2}\frac{ \left<\Theta_\ell^{(m)}(\vek{k})\,\dot\Theta_{\ell}^{(m)}(\vek{k}')\right>}{(2\ell+1)}
\nonumber\\
&=\int \!\frac{\id^3 k}{(2\pi)^3} \id^3 k'\,\expf{i \scriptsize \vek{x}\cdot(\vek{k}-\vek{k}')}
\sum_{\ell} \!\sum_{m=-2}^{m=2}\frac{\Theta_\ell^{(m)}(k)\,\dot\Theta_{\ell}^{(m)}(k') \,\delta(\vek{k}-\vek{k}') P^{(m)}_i(k) }{(2\ell+1)}
\nonumber\\
\label{eq:TI2_term}
&=\sum_{\ell} \!\sum_{m=-2}^{m=2}\int \!\frac{k^2 \! \id k}{2\pi^2} P^{(m)}_i(k) 
\frac{\Theta_{\ell}^{(m)}(k)\,\dot \Theta_{\ell}^{(m)}(k)}{(2\ell+1)}.
\end{align}
For $\dot\Theta_I$ we use the scattering terms given in \citet{Hu1997}. Notice that we do not use a different variable to denote the transfer functions, but whenever a power spectrum appears explicitly, this is what is meant.

To evaluate $\!\id \big<\Theta_+ \Theta_-\big>/\!\id t$, we follow very similar steps. Again, for the time derivatives, we only need to account for the scattering terms. We start with the angle average of $\Theta_+\dot \Theta_-=\Theta_+\dot \Theta^\ast_+$
\beal
\int \Theta_+\dot \Theta^\ast_+ \frac{\id^2 n}{4\pi}&=\int \!\frac{\id^3 k}{(2\pi)^3}\frac{\id^3 k'}{(2\pi)^3}\,\expf{i \scriptsize \vek{x}\cdot(\vek{k}-\vek{k}')}
\!\sum_{\ell,\ell'} \!\sum_{m, m'} 
\frac{(-i)^{\ell}i^{\ell'} }{\sqrt{(2\ell+1)(2\ell'+1)}} 
\nonumber\\
&\qquad\times\Theta_{+,\ell}^{(m)}(\vek{k})\,[\dot\Theta_{+,\ell'}^{(m')}(\vek{k}')]^\ast
\int {}_2 Y_{\ell m}(\vek{n}) \,{}_2 Y^\ast_{\ell' m'}(\vek{n}) \id^2 n 
\nonumber\\
&=\int \!\frac{\id^3 k}{(2\pi)^3}\frac{\id^3 k'}{(2\pi)^3}\,\expf{i \scriptsize \vek{x}\cdot(\vek{k}-\vek{k}')}
\!\sum_{\ell,\ell'} \!\sum_{m, m'} 
\frac{(-i)^{\ell}i^{\ell'} }{\sqrt{(2\ell+1)(2\ell'+1)}} 
\nonumber\\
&\qquad\times\Theta_{+,\ell}^{(m)}(\vek{k})\,[\dot\Theta_{+,\ell'}^{(m')}(\vek{k}')]^\ast  \delta_{\ell,\ell'}\delta_{m,m'}
\nonumber\\ \nonumber
&=\int \!\frac{\id^3 k}{(2\pi)^3}\frac{\id^3 k'}{(2\pi)^3}\,\expf{i \scriptsize \vek{x}\cdot(\vek{k}-\vek{k}')}
\sum_{\ell} \!\sum_{m=-2}^{m=2}\frac{\Theta_{+,\ell}^{(m)}(\vek{k})\,[\dot\Theta_{+,\ell}^{(m)}(\vek{k}')]^\ast}{(2\ell+1)}.
\end{align}
The ensemble average with $\big<\delta^{(m)}(\vek{k})\,[\delta^{(m)}(\vek{k}')]^\ast\big>=(2\pi)^3 \delta(\vek{k}-\vek{k}') P^{(m)}_i(k)$ then simplifies to
\beal
\big<\Theta_+\dot \Theta_-\big>&
=\int \!\frac{\id^3 k}{(2\pi)^3}\frac{\id^3 k'}{(2\pi)^3}\,\expf{i \scriptsize \vek{x}\cdot(\vek{k}-\vek{k}')}
\sum_{\ell} \!\sum_{m=-2}^{m=2}\frac{\left<\Theta_{+,\ell}^{(m)}(\vek{k})\,[\dot\Theta_{+,\ell}^{(m)}(\vek{k}')]^\ast\right>}{(2\ell+1)}
\nonumber\\ \nonumber
&=\sum_{\ell} \!\sum_{m=-2}^{m=2}\int \!\frac{k^2\! \id k}{2\pi^2} P^{(m)}_i(k) 
\frac{\Theta_{+,\ell}^{(m)}(k)\,[\dot\Theta_{+,\ell}^{(m)}(k)]^\ast}{(2\ell+1)}.
\end{align}
For $\big<\Theta_-\dot \Theta_+\big>$, the steps are similar, so that finally we find
\beal
\label{eq:Tpm2_term}
\frac{\id\left<\Theta_+ \Theta_-\right>}{\id t}
&=2\sum_{\ell} \!\sum_{m=-2}^{m=2}\int \!\frac{k^2\!\id k}{2\pi^2} P^{(m)}_i(k) 
\frac{E^{(m)}_\ell\dot E^{(m)}_\ell+B^{(m)}_\ell\dot B^{(m)}_\ell}{(2\ell+1)},
\end{align}
where for the last step we used the definition $\Theta^{(m)}_{\pm,\ell}=E^{(m)}_{\ell}\pm i B^{(m)}_{\ell}$ and that the transfer functions  $E^{(m)}_{\ell}$ and $B^{(m)}_{\ell}$ are real. The final expression for the heating integral can then be obtained by inserting the expressions for the collision term from \citet{Hu1997}.

\vspace{-4mm}
\section{Evolution of tensor amplitude}
\label{app:tensor}
Using $a'/a = \mathcal{H}$ ($X'\equiv \partial X/\partial \eta$), $R_\nu=\rho_\nu/(\rho_\gamma+\rho_\nu)\approx 0.41$ and the Friedmann equation during radiation domination $\mathcal{H}^2\approx 8\pi G a^2 (\rho_\gamma+\rho_\nu)/3$, the equation of motion for the amplitude of tensor perturbations, $h$, can be written as \citep{Hu1997}:
\beal\label{eq:tens_ev}
h'' \!\!+\!2 \mathcal{H} h' \!\!+ \!k^2 h 
&\!=\! 8\pi G a^2 \!\left[ p_\gamma \pi^{(2)}_\gamma\! + \!p_\nu \pi^{(2)}_\nu\right]
\!=  \!\mathcal{H}^2 \!\left[ (1-R_\nu)\pi^{(2)}_\gamma \!+\! R_\nu \pi^{(2)}_\nu\right]\!.
\end{align}
Here, $\pi^{(2)}_i$ are the contribution to the anisotropic stress from photons and neutrinos. We also have $\eta=\int c \id t /a \propto a$, and thus $\mathcal{H}=\eta^{-1}$. Following \citet{Hu1997}, we used the convention $h_{ij}=2 h Q^{(2)}_{ij}$ for the tensor perturbations. Here, $Q^{(2)}_{ij}$ are Laplacian eigenfunctions $\nabla^2 Q^{(\pm 2)}_{ij}=-k^2 Q^{(\pm 2)}_{ij}$ with explicit representation $Q^{(\pm 2)}_{ij}=\sqrt{3/8}\,(\vek{e}_1\pm i \vek{e}_2)_i \otimes(\vek{e}_1\pm i \vek{e}_2)_j \,\expf{i \vek{k}\cdot\vek{x}}$.

\subsection{Free evolution of the tensor amplitude}
Neglecting anisotropic stress directly gives the free solution for tensors
\beal
h=A(k) \frac{\sin k \eta}{k \eta} + B(k) \frac{\cos k \eta}{k \eta}.
\end{align}
From the initial condition $h' = 0$ as $\eta \rightarrow 0$, we need $B=0$, so that the undamped solution is 
\bsub
\beal
h_{\rm free}&=A(k)\frac{\sin k \eta}{k\eta}
\\
h'_{\rm free}&= \frac{h_{\rm free}}{\eta} [k \eta \cot k \eta - 1] = \frac{A(k)}{\eta} \left[\cos k \eta -\frac{\sin k \eta}{k\eta}\right].
\end{align}
\esub
This solution has no characteristic scale at which perturbations cut off at small scales, however, the initial conditions introduce a small-scale cutoff related to the end of inflation, $k_{\rm end}$, and reheating \citep[e.g.,][]{Boyle2008a, Watanabe2006}.

\subsection{Evolution of tensor amplitude with damping by photons}
\label{app:gamma_damp}
Neglecting neutrinos, with $\pi^{(2)}_\gamma=(8/5)\Theta^{(2)}_2$ \citep{Hu1997}, we have
\beal
h'' \!+\!2h'/\eta  \!+\! k^2 h 
&\!\approx \!\mathcal{H}^2 \frac{8}{5} \Theta^{(2)}_2(1-R_\nu)
\!\approx \!- \mathcal{H}^2 \frac{32(1-R_\nu)}{15\tau'} h'
=-\Gamma_\gamma h',
\end{align}
where in the last step we used $\Theta^{(2)}_2\approx -(4/3) (h' /\tau')$ from the tight coupling solution. Transfer effects modify the r.h.s of this equation [see Eq.~\eqref{eq:damping_coefficient}], but the corrections are energetically not crucial for the evolution of $h$.

To include damping due to photons, we use $\tau'\propto \eta^{-2}$, so that during radiation domination we have $\Gamma_\gamma=32 \mathcal{H}^2 (1-R_\nu)/[15\tau' ]\approx {\rm const}$. The damped solution for initial condition $h' = 0$ as $\eta \rightarrow 0$ therefore reads
\beal
h=A(k)\expf{ - i k \eta (1+\xi )- \frac{\Gamma_\gamma}{2}\eta }
{}_1F_1\left(1+\Gamma_\gamma/(2ik [1+\xi]) , 2, 2 i k \eta[1+\xi]\right)
\end{align}
with $\xi=\sqrt{1-[\Gamma_\gamma/(2k)]^2}-1\approx \mathcal{O}(\Gamma_\gamma^2/k^2)$. We can rewrite $\Gamma_\gamma$ in terms of the standard photon damping scale, $\kD$. Comparing with $\partial_\eta \kD^{-2}=8/[45\tau']\propto \eta^2$ (neglecting baryon loading), we thus have $\Gamma_\gamma\approx 12 \mathcal{H}^2(1-R_\nu)\partial_\eta \kD^{-2}$, which with $\partial_\eta \kD^{-2}=3/[\eta \kD^2]$ gives
\beal
\Gamma_\gamma\eta \approx \frac{36(1-R_\nu)}{(\eta \kD)^2}=36 \frac{\mathcal{H}^2(1-R_\nu)}{\kD^2}
\simeq 10 a (1-R_\nu), 
\end{align}
where we used $\eta \kD \simeq 1.9/\sqrt{a}$. At $z\gtrsim 10^4$, we thus have $\Gamma_\gamma \eta \lesssim 10^{-3}$. Restricting ourselves to small scales (say $k\gtrsim 0.01 \Mpc^{-1}$), we find
\beal
h\approx h_{\rm free}(k, \eta)\, \expf{-\frac{\Gamma_\gamma\eta}{2}},
\end{align}
which describes the damping of the tensor mode amplitude due to anisotropic stress from photons. Overall, this is a tiny correction to the total energy density of gravity waves, and thus usually can be neglected.

\vspace{0mm}
\subsection{Evolution of tensor amplitude with damping by neutrinos}
The anisotropic stress contributed by massless neutrinos was derived in \citet{Weinberg2004} and takes the form
\beal
\pi^{(2)}_\nu &=  - 24 \int_0^\eta K(k [\eta-\eta^*]) \,h' (\eta^*) \!\id\eta^*
\nonumber\\
K(x)&
=\frac{1}{16}\int_{-1}^1 (1-s^2)^2 \expf{i s x} \id s 
= \frac{3\sin x}{x^5}-\frac{3\cos x}{x^4} -\frac{\sin x}{x^3}.
\end{align}
Using this expression, one can numerically solve Eq.~\eqref{eq:tens_ev}. The overall effect is that at very small scales the amplitude of the tensor perturbations is reduced to $A_{\rm damp}\simeq 0.8 A$. This effect can be captured analytically using spherical Bessel functions \citep{Dicus2005}
\beal
h(k, \eta)&\approx A(k) \sum_n a_n j_n(k\eta) 
\nonumber \\
\label{eq:sol_h_damping}
h'(k, \eta)&\approx \frac{A(k)}{\eta} \sum_n a_n [n j_{n}(k\eta)-k\eta j_{n+1}(k\eta) ] 
\end{align}
with $a_0=1$, $a_2=0.243807$, $a_4=\pot{5.28424}{-2}$ and $a_6=\pot{6.13545}{-3}$. With this expression, we can directly compute the tensor contribution to the heating rate of the photon field.

\vspace{-0mm}
\section{Approximate solutions for the photon transfer functions}
\label{app:ana_sol_env}
Although in the free streaming phase one does expect higher multipoles to become significant, 
our numerical analysis shows that the main features of the solution can be captured already when only including multipoles for $\ell=2$. The Boltzmann hierarchy for this case reads
\bsub
\beal
\partial_\eta \Theta^{(2)}_2&=-\tau' \left(\frac{9}{10}\Theta^{(2)}_2+\frac{\sqrt{6}}{10}E^{(2)}_2\right) - h'
\\
\partial_\eta E^{(2)}_2&=-\tau' \left(\frac{2}{5}E^{(2)}_2+\frac{\sqrt{6}}{10}\Theta^{(2)}_2\right) - k\frac{2}{3}B^{(2)}_2
\\
\partial_\eta B^{(2)}_2&=-\tau' B^{(2)}_2 + k\frac{2}{3}E^{(2)}_2.
\end{align}
\esub
The perturbations are sourced by $h'$ in the equation for $\Theta^{(2)}_2$. Without scattering neither $E^{(2)}_2$ or $B^{(2)}_2$ would be excited. For $k\ll \tau'$ and under quasi-stationary conditions (no time derivatives), we can readily verify the tight coupling approximations, $B^{(2)}_2\approx 0$, $\frac{\sqrt{6}}{10}\Theta^{(2)}_2\approx -\frac{2}{5}E^{(2)}_2\Rightarrow E^{(2)}_2\approx -\frac{\sqrt{6}}{4}\Theta^{(2)}_2$ and thus $\Theta^{(2)}_2\approx -(4/3)\,h'/\tau'$.

The system behaves like a driven coupled oscillator in all relevant regimes. The amplitudes of the individual components depend on the tightness of the coupling terms mediated by Thomson scattering. In the regime $\xi =k/\tau' \ll 1$, all components follow suit with the driving force, while for $\xi\gg 1$, phase shifts develop and the oscillation amplitudes decay. Making the ansatz $\Theta^{(2)}_2=A_\Theta \expf{i k\eta}$, $E^{(2)}_2=A_E \expf{i k\eta}$ and $B^{(2)}_2=A_B \expf{i k\eta}$, for a driving force $h' = A_h \expf{i k\eta}$, the fastest variation of the solutions is captured by $\simeq \expf{i k\eta}$, while the variations of the phases and amplitudes are slow over time-scales $\simeq 1/k$. Putting things together, we thus find
\bsub
\beal
i k A_\Theta&=-\tau' \left(\frac{9}{10} A_\Theta +\frac{\sqrt{6}}{10}A_E \right) - A_h
\\
i k A_E&=-\tau' \left(\frac{2}{5}A_E+\frac{\sqrt{6}}{10}A_\Theta\right) - k\frac{2}{3}A_B
\\
i k A_B&=-\tau' A_B + k\frac{2}{3}A_E.
\end{align}
\esub
The solutions for the amplitudes read
\bsub
\label{eq:appro_transfer_functions}
\beal
\label{eq:appro_transfer_functions_a}
\frac{|A_\Theta|}{\frac{4}{3}\,\frac{|A_h|}{\tau'}}&=  \sqrt{\frac{1+\frac{341}{36}\xi^2 + \frac{625}{324}\xi^4}{1+\frac{142}{9}\xi^2 + \frac{1649}{81}\xi^4 + \frac{2500}{729} \xi^6}}
\\
\frac{|A_E|}{\frac{4}{3}\,\frac{|A_h|}{\tau'}}&= \frac{\sqrt{6}}{4}
\sqrt{\frac{1+\xi^2}{1+\frac{142}{9}\xi^2 + \frac{1649}{81}\xi^4 + \frac{2500}{729} \xi^6}}
\\
\frac{|A_B|}{\frac{4}{3}\,\frac{|A_h|}{\tau'}}
&= \frac{\xi}{\sqrt{6}}\,\sqrt{\frac{1}{1+\frac{142}{9}\xi^2 + \frac{1649}{81}\xi^4 + \frac{2500}{729} \xi^6}}.
\end{align}
These expressions show that in the free streaming regime ($\xi\gg 1$), the amplitude of $\Theta_2^{(2)}$ drops as $\propto \xi^{-1}$, while for $E_2^{(2)}$ and $B_2^{(2)}$ one finds a faster decay, $\propto \xi^{-2}$.
By evaluating these expressions and taking the low/high-$\xi$ limits, it is easy to verify that
$\id \ln |A_{\Theta,E,B}|/\id \eta\ll k$, confirming the approximation made above.
For the phase relation, we find
\beal
\label{eq:phases_a}
\tan\phi_\Theta &= -\frac{11}{6} \xi\,
\frac{1 +\frac{697}{99} \xi^2+\frac{1250}{891} \xi^4}{1 +\frac{197}{18} \xi^2+\frac{125}{54} \xi^4} 
\\
\label{eq:phases_b}
\tan (\phi_E-\pi)&= -\frac{13}{3} \xi\,\frac{1 +\frac{121}{117} \xi^2}{1 - \xi^2- \frac{50}{27}\xi^4} 
\\
\tan(\phi_B-\pi)&= -\frac{16}{3} \xi\,\frac{1 - \frac{25}{72} \xi^2}{1 - \frac{19}{3}\xi^2}.
\end{align}
\esub
In the tight coupling regime, the phases all vanish and the photons follow suit with the driving force, although both $E^{(2)}_2$ and $B^{(2)}_2$ start with the opposite sign of $\Theta^{(2)}_2$. For large $\xi$, both $\Theta_2^{(2)}$ and $B_2^{(2)}$ are $\pi/2$ out of phase with the driving force, while $E_2^{(2)}$ is locked in phase.

\vspace{-0mm}
\subsection{Slightly higher precision}
\label{app:high_prec}
Our numerical results show that the largest correction beyond $\ell=2$ is captured by adding terms with $\ell=3$. Here, we only consider $\Theta_2^{(2)}$ and $\Theta_3^{(2)}$. Proceeding like for $\ell=2$, we find
\beal
\frac{|A_{\Theta_2}|}{\frac{4}{3}\frac{|A_h|}{\tau'}}&\approx  
\sqrt{\frac{1+13.7 \xi^2+39.1
   \xi^4+41.1 \xi^6+14.8
   \xi^8+ 0.170 \xi^{10}}{1+20.5 \xi ^2+85.5
   \xi ^4+136 \xi ^6+89.7
   \xi ^8+19.8 \xi ^{10}+0.222 \xi^{12}}}
\nonumber\\[-0.5mm] \nonumber
\frac{|A_{\Theta_3}|}{\frac{4}{3}\frac{|A_h|}{\tau'}}&\approx  
\sqrt{ \frac{(\xi^2/5)[1+12.7 \xi
   ^2+26.5 \xi
   ^4+14.6 \xi ^6+0.170
   \xi ^8]}{1+20.5 \xi ^2+85.5
   \xi ^4+136 \xi ^6+89.7
   \xi ^8+19.8 \xi ^{10}+0.222 \xi^{12}}}
\end{align}
to reproduce the full numerical result well. The coefficients were obtained by evaluating the ratios that appeared when solving the algebraic system. With this, we find the improved transfer function
\beal
\label{eq:appro_transfer_functions_BB_sum}
\mathcal{T}_\Theta
&\approx 
\frac{1+13.9 \xi^2+41.6 \xi^4+46.1 \xi^6+17.6 \xi^8+0.203 \xi^{10}}
{1+20.5 \xi^2+85.5 \xi^4+136 \xi^6+89.7\xi ^8+19.8 \xi^{10}+0.222 \xi^{12}},
\end{align}
which reproduces the full numerical result to at higher precision. Still the overall correction remains small, unless the power spectrum is very blue.

\end{appendix}

\bibliographystyle{mn2e}
\bibliography{Lit}

\end{document}